\documentclass[reprint,
    preprintnumbers,
    superscriptaddress,
    nofootinbib,
     amsmath,amssymb,
     aps,
     prd,
    floatfix,
longbibliography
]{revtex4-2}

\usepackage[paperwidth=210mm,paperheight=297mm,hmargin=1.3cm,vmargin=2.5cm]{geometry}
\usepackage{verbatim}
\usepackage{graphicx}
\usepackage{dcolumn}
\usepackage{bm}
\usepackage{xcolor}
\usepackage{pifont}
\usepackage{hyperref}
\hypersetup{
  colorlinks   = true, 
  urlcolor     = blue, 
  linkcolor    = blue, 
  citecolor   = red 
}

\usepackage{subcaption}
\usepackage{lipsum}
\usepackage{ragged2e}

\begin{document}

\title{Bunch-Davies initial conditions and non-perturbative inflationary dynamics \\ in Numerical Relativity}

\author{Yoann L. Launay}
 
  \email{yoann.launay@outlook.com}
 \affiliation{Centre for Theoretical Cosmology, Department of Applied Mathematics and Theoretical Physics,
University of Cambridge, Wilberforce Road, Cambridge CB3 0WA, United Kingdom}

\author{Gerasimos I. Rigopoulos}
\email{gerasimos.rigopoulos@ncl.ac.uk}
\affiliation{School of Mathematics, Statistics and Physics,  Newcastle University, Newcastle upon Tyne, NE1 7RU, United Kingdom}

\author{E. Paul S. Shellard}
\email{eps1@cam.ac.uk}
\affiliation{Centre for Theoretical Cosmology, Department of Applied Mathematics and Theoretical Physics,
University of Cambridge, Wilberforce Road, Cambridge CB3 0WA, United Kingdom}

\begin{abstract}
We show that it is possible to simulate realistic inhomogeneities during cosmological inflation with high precision using {Numerical Relativity}. Stochastic initial conditions are set in line with the Bunch-Davies {vacuum} and satisfy the Hamiltonian and Momentum constraints of General Relativity to leading order in perturbation theory. The subsequent fully non-linear dynamical evolution is formulated within a family of geodesic gauges but can in principle be adapted to any choice of coordinates. We present 3 examples of inflationary dynamics: a simple quadratic potential, a potential with an inflection point and a strong resonance model. When perturbations are small, we recover standard predictions of cosmological perturbation theory, and we quantify strongly non-linear inhomogeneities when non-perturbative configurations emerge, such as in the strong resonance model. Our results pave the way towards the first realistic non-perturbative, and fully non-linear Numerical Relativity simulations of the early inflationary universe. 
\end{abstract}

\maketitle

\section{Introduction \label{sec:intro}}
The theory of inflation is arguably the best contemporary model available to explain the observed inhomogeneities in the temperature of the Cosmic Microwave Background (CMB) \cite{PlanckInflation}. It describes a period in the history of the universe before any of today's fields and particles existed, including those present at CMB times. Instead, the proposed inflaton field played the leading role, being subject to quantum fluctuations that grow and freeze into the spacetime geometry during the fastest known epoch of accelerating expansion. 
Similarly to the achievements in flat space for particle physics, cosmologists have developed a framework for Quantum Field Theory on curved spacetime (QFTCS) to predict spacetime correlators \cite{WaldQFTCS}. Small spacetime inhomogeneities in spacetime carry correlations as a result of the shared vacuum initial condition in the far past and their subsequent interactions and such correlations may be observable today.

Within the inflationary paradigm, analytical predictions for different models and, more recently, classes of models have been calculated for higher-order correlators over the past two decades \cite{Maldacena_2003, chen_primordial_2010, arkani-hamed_cosmological_2020,wang_bootstrapping_2022, DuasoPueyo23}. They all share a key restriction: the assumption of perturbativity. Model-independent techniques often rely on simplifications afforded by assumed symmetries. In particular, loop computations in cosmologically relevant spacetimes are far less comprehensive than those of flat spacetime due to the analytical challenges they pose \cite{Senatore10,Assassi12}.

To move beyond simple models and symmetries, numerical methods must be employed. The leading Fourier amplitude of quantum fluctuations can be obtained from the Mukhanov-Sasaki equations for the perturbations \cite{Mukhanov1981,Mukhanov92}
given (numerical) solutions of the background Friedmann equations for inflation. 
For the computation of higher-order correlation functions of quantum cosmological perturbations several methods have been developed, among which are  the \textit{in-in} formalism computations \cite{ Maldacena_2003, chen_large_2007,clarke_probing_2021} together with transport equations \cite{mulryne_pytransport_2017, werth23}. So far, these have been implemented numerically mostly up to the bispectrum and for cubic order interactions, though in theory they are not limited to these. However, as mentioned above, the absence of a general treatment of quantum loops on a curved spacetime implies, despite the existence of some analytical examples and methods \cite{Senatore10, Assassi12, lee_leading_23, Iacconi24}, that none of the mentioned numerical techniques seem to tackle generic loop computations. In principle, some hope can be found using flow equations up to a certain order
\cite{dias_numerical_2016,werth23} but the computational resources needed would increase drastically together with the difficulty of the task. 

However, for many interesting and realistic models, these effects are suppressed in the long-wavelength limit when crossing the Hubble radius \cite{Weinberg05,Weinberg06, burgess_minimal_2022}.  This is noteworthy because it makes it feasible to study the evolution of the relativistic and classical equations of motion alongside non-perturbative scenarios. In general, these equations can fully account for classical non-linearity but also for all orders in perturbation theory up to numerical precision, the studied range of scales, and simplifying assumptions. From this context emerged the separate universe approach \cite{wands_new_2000, salopek_nonlinear_1990, rigopoulos_separate_2003, Rigopoulos_multi06} and
the $\delta N$ approach \cite{wands_new_2000, Lyth03} but also the stochastic inflation community from great analytical use of the super-Hubble limit \cite{Starobinsky_stochastic_1988, morikawa_dissipation_1990, salopek_stochastic_1991, vennin_correlation_2015} together with multiple recent codes \cite{Jackson22,cruces_stochastic_2022,tomberg_numerical_2023,Mizuguchi24,Animali25}. Recent advances \cite{Jackson22, cruces_stochastic_2022, Launay24} have demonstrated that this limit and the accompanying analytical simplifications do not, however, describe all scenarios because the long-wavelength approximation neglects gradient terms and associated key phenomena.

Cosmological computations on a lattice have recently emerged as an alternative computational approach that can include spatial gradients \cite{latticeeasy,Easther10,Caravano22thesis,Cosmolattice}, which in return neglect the spacetime geometry, and the stochastic noise at each time step, thus unable to provide a rigourous coarse-graining of quantum diffusion. Recent work \cite{Caravano2024} has presented the first non-perturbative non-separate universes simulations of a field from a box initialized with full Gaussian initial conditions.
Although these recent advances incorporate non-perturbative global backreaction regimes or local perturbative ones \cite{caravano2024axionbackreaction,caravano2024usr}, the full geometric dynamics and associated subtleties (full backreaction, coordinate choice, etc) are not accessible.

This indicates that Numerical Relativity (NR), or full General Relativity on a lattice, is necessary. It is worth noting that inflationary spacetimes have recently been successfully simulated using such technologies, building on the expertise and challenges encountered in black hole simulations - see e.g.~\cite{aurrekoetxea24} for a review of cosmology using NR. The focus has been on post-inflationary evolution and preheating dynamics, but also on inflationary spacetimes, focusing on the study of the robustness of inflation given generic initial conditions \cite{East16,clough_robustness_2017,Bloomfield19,joana_inhomogeneous_2021, aurrekoetxea_effects_2020, Corman23,Elley2024}. 

As in other physical setups, the primary difficulty in employing NR for cosmology has been the construction of initial conditions which must satisfy the highly non-linear constraints of general relativity. Starting from simplified analytical solutions \cite{Clough_2015}, this community made great use of numerical solvers for the constraints to provide initial inhomogeneities in the inflaton and its momentum together with a 3D geometry \cite{East12,aurrekoetxea_cttk_2022}. Because of its assumptions such as conformal flatness, these methods are only good if the exact form of the perturbations does not matter \cite{aurrekoetxea24}. Previous applications include the study of inflationary exit scenarios, such as recent preheating studies \cite{giblin_jr_preheating_2019,aurrekoetxea_oscillon_2023}.

Beyond exploring the more exotic scenarios of inflation and deepening our understanding of inflationary spacetimes, the next step for these simulations seems to be moving from qualitative and convenient simplifications to quantitative analysis that provides high-precision predictions for direct comparison against observation, including present and future CMB and large-scale structure surveys. To achieve this, NR still needs to start from the exact initial conditions provided by QFTCS, such as those sourced by the Bunch-Davies vacuum, which involve a complete spectrum. By doing so, we can move beyond previous studies, accounting for all non-linearities, non-perturbative effects, and a broader range of models, as we will explore here.

This article's main results are presented in the three following sections: Section \ref{sec:startingspacetime} introduces the initial value problem and in particular a linear solution for the entire initial slice for scalar perturbations for a chosen numerical gauge, adapting previous work to the NR problem \cite{Launay24}. Section \ref{sec:nummethods} showcases the full numerics pipeline and applies it on a vanilla slow-roll type of inflation, using the stochastic initial conditions introduced in the previous section as the starting point of a full GR simulation. As perturbations remain small in this case, linear perturbation theory results are fully recovered from our simulations, thereby offering a useful validation check. Section \ref{sec:Applications} is a proof of concept that fully non-perturbative dynamics of the scalar field and the geometry can be probed. We showcase this by studying  inhomogeneous inflation with an inflection point potential that exhibits ultra slow-roll, and a strong resonance model leading to spatial regions of eternal inflation. We conclude in section \ref{sec:sec5}.  Appendices \ref{app:dicBSSN} and \ref{app:extract} provide useful conversions between cosmological perturbation theory and evolved NR quantities, the evolution equations of which are provided in Appendix \ref{app:BSSN}. Appendix \ref{app:random} explains our stochastic generation method on the initial lattice, while Appendix \ref{app:RNL} defines the non-perturbative generalisation of the curvature perturbation on comoving hypersurfaces. Appendix \ref{app:UVdep} shows why deep UV modes cannot be investigated in this framework. 
\paragraph*{Notations.} $\Delta = \partial_i \partial^i$. Bold symbols are random variables and hats designate quantum operators. $M_{Pl}$ is the reduced Planck mass. The inflaton rolls from left to right as a convention.
\section{Starting a spacetime \label{sec:startingspacetime}}
In this section, we lay the foundations behind our new \textit{python} code {STO}chastic {I}nflation {I}nitial {C}onditions for {GR} (STOIIC-GR), soon to be made open-access.

\subsection{The initial value problem \label{subsec:initValprob}}
In this subsection, we introduce useful notations while reminding some basics of general relativity not commonly emphasized in a cosmology context, pointing out where the difficulty arises in solving. An expert reader might wish to proceed to the next subsection.

In general relativity, a 3+1 spacetime is described by a metric of the form
\begin{equation}
    d s^2=-\alpha^2 d t^2+\gamma_{i j}\left(d x^i-\beta^i d t\right)\left(d x^j -\beta^j d t\right).
    \label{eq:ADMmetric}
\end{equation}
When coupled to a scalar field $\phi$ with conjugate momentum $\Pi$
\begin{equation}
\Pi=\frac{1}{\alpha}\left(\dot{\phi}+\beta^i \partial_i \phi\right),
    \label{eq:Mom}
\end{equation}
the system evolves according to a set of partial differential equations such as the ADM equations \cite{ADM}
\begin{equation}
\left \{
\begin{aligned}
\frac{1}{\alpha}\left(\dot{\Pi}+\beta^i \Pi_{\mid i}\right)-K \Pi-\frac{\alpha^{\mid i}}{\alpha} \phi_{\mid i}-\phi_{\mid i}^{\mid i}+\frac{d V}{d \phi}=0,&\\
 \dot{K}+\beta^i K_{, i}+\alpha^{\mid i}{ }_{\mid i}-\alpha\left({ }^{3} R+K^2\right)&\\
 - M_{Pl}^{-2} \alpha\left(\frac{1}{2} S-\frac{3}{2} \rho\right) = 0, & \\
  \dot{\tilde{K}}_{ij}+2\alpha\tilde{K}_{il}{\tilde{K}}^l{ }_j+\beta^k \tilde{K}_{ij \mid k}-2\beta_i{ }^{\mid k} \tilde{K}_{jk}+ \alpha_{\mid i\mid j}&\\
  -\frac{1}{3} \alpha^{\mid k}{ }_{\mid k} \delta_{ij} 
-\alpha\left({ }^3\tilde{R}_{ij}+\frac{1}{3}K \tilde{K}_{ij}\right) + M_{Pl}^{-2} \alpha \tilde{S}_{ij}=0, &
\end{aligned}\right .
    \label{eq:GRADM}
\end{equation}
where we have defined the Ricci tensor ${}^3R_{ij}$ of the spatial metric $\gamma_{ij}$, decomposed into the Ricci scalar ${}^3R$ and the traceless ${}^3\tilde{R}_{ij}$, and the extrinsic curvature $K_{ij}$ with respect to $\gamma_{ij}$, 
\begin{equation}
K_{ij} \equiv  -\frac{1}{2\alpha}(\gamma_{i j,0}+\beta_{i|j}+\beta_{j|i}),
    \label{eq:eqK}
\end{equation}
with $K$ and $\tilde{K}_{ij}$ being its trace and traceless parts accordingly,
and the components of the energy-momentum tensor, the energy and momentum densities
\begin{equation}
\left\{
\begin{aligned}
\rho & =\frac{1}{2} \Pi^2+\frac{1}{2} \partial_i \phi \partial^i \phi+V(\phi), \\
{\cal { J}}_i &= -\Pi \partial_i\phi,
\end{aligned}\right .
\label{eq:densities}
\end{equation}
and the stress tensor $S_{i j} = T_{ij}$, decomposed as
\begin{equation}
    \left \{
    \begin{aligned}
        S & =\frac{3}{2} \Pi^2-\frac{1}{2} \partial_k \phi \partial^k \phi-3 V(\phi), \\
\tilde{S}_{i j}& =  \frac{1}{2}\left(\partial_i \phi \partial_j \phi-\frac{1}{3} \partial_k \phi \partial^k \phi \,\delta_{i j}\right) .
    \end{aligned} \right .
        \label{eq:SETdef}
\end{equation}
{A vertical bar denotes a covariant derivative associated with $\gamma_{ij}$}. Given this system, evolution could be naively started from initial data for the quantities $(\gamma_{ij}, K_{ij}, \phi, \Pi)$ on a spatial hypersurface $\Sigma$. 

However, such a system is under-determined because of the gauge freedom left: each or most of the previous quantities {involve a $3+1$ split of tensor components} and so basis-dependent objects. For that reason, the coordinates need to be specified by four algebraic constraints in a four-dimensional spacetime. As an example, one could impose the {reference frame} of a stationary observer with respect to a source. Writing the evolution equations in the ADM form \eqref{eq:GRADM} is more convenient in such scenarios as the non-dynamical lapse $\alpha$ and the shift $\beta^i$ can be evolved to impose the coordinate system we want. 

In any case, the physical result also needs to be independent of the coordinate system, which is why the tensor fields need to be solutions of the so-called \textit{Hamiltonian} and \textit{Momentum} constraints
\begin{equation}
\left \{
\begin{aligned}
{\cal H} & \equiv  { }^{3}R+\frac{2}{3} K^2-\tilde{K}_{i j} \tilde{K}^{i j}-2 M_{Pl}^{-2} \rho = 0, \\
{\cal M}_i & \equiv  \tilde{K}^j{}_{i \mid j}-\frac{2}{3} K_{\mid i}- M_{Pl}^{-2} {\cal { J}}_i=0. 
\end{aligned}\right .
    \label{eq:cons_eq}
\end{equation}
Indeed, if these are satisfied, any coordinate change will leave the action invariant (${\cal H}$ and ${\cal M}_i$ are the generators of these transformations). In doing so, we are also making sure that our spacetime is physical by ensuring the balance between the curvature's and the contents' energy and momentum densities. 

These equations are highly non-linear and must be satisfied at all times over the entire spatial hypersurfaces. In particular, they crucially need to be satisfied at the initial spatial slice. Both analytical and numerical solutions usually require assumptions on the fields' symmetries for feasibility \cite{baumgarte_shapiro_2010}, especially for generic distributions of the spacetime's energy-momentum content. For these reasons, it is often said that the real problem to solve in GR is the initial value problem. Once initially satisfied, the constraints are then preserved by the evolution equations, although this property sometimes needs to be enforced in numerical evolution schemes (see constraints damping \cite{CCZ41, CCZ42}), to counter numerical error building up at each time step. 

\subsection{Satisfying the inhomogeneous constraints}\label{subsec:Inhom-Constr}
The constraints have been studied in the case of inhomogeneous and inflationary spacetimes. A first analytical solution utilized simplified and hypothetical perturbations in the scalar field \cite{clough_scalar_2017, macpherson_inhomogeneous_2017}. Given the heuristics of the methods, it became quickly urgent to solve for more complex perturbations, hence the need for numerical solvers \cite{aurrekoetxea_cttk_2022, GRTresna}, building from the knowledge of the black hole numerics community. In these most recent numerical works, initial inhomogeneous perturbations of the extrinsic curvature are solved for given perturbations of the field and its momentum on an initially flat metric (conformal flatness together with $a = 1$ at the start). The constraints are satisfied up to a very good accuracy and are not limited to perturbative regimes. 

It is however important to remember that an admissible solution with respect to the constraints is not necessarily a physical one. In particular, any of these works could encounter opposition regarding the physical relevance of the assumptions and approximations made. For instance, it is not obvious how much of the simplifications are part of the gauge freedom and thus authorised.

As suggested in \cite{giblin_jr_preheating_2019}, it is however possible to solve the constraints perturbatively, trading the initial non-linearity for a more physically-motivated and analytically controlled starting point. This was studied more extensively in \cite{Launay24} for any gauge but in another context. We will use some of the derivations below. Note that the present work differs at least in using a time-independent window function of the spectra.

Indeed, a reasonable assumption that can be made for an inflationary spacetime is that of a metric close to FLRW's
\begin{equation}
\begin{aligned}
    d s^2=-\alpha_b^2(1+2 \Psi) d t^2+2 a^2 B_{, i} d t d x^i & \\
    +a^2\left[(1-2 \Phi) \delta_{i j}+2 E_{, i j}\right] d x^i d x^j,
\end{aligned}    
    \label{eq:SVTmetric}
\end{equation}
where the four functions $\Phi, \Psi, B$ and $E$ define the scalar perturbations and $a$ the usual background scale factor. One can verify that at 0th order, noted with subscript ${}_b$, the Hamiltonian constraint and the 
evolution equations for the extrinsic curvature and the field  yield the homogeneous and isotropic Friedmann and field equations of inflation
\begin{equation}
\left\{\begin{aligned}
6 H^2 - 2M_{Pl}^{-2}\left(V(\phi_b)+\frac{1}{2}\frac{\dot{\phi_b}^2}{\alpha_b^2}\right) & =  0, \\
3M_{Pl}^{-2}V(\phi_b)-9 H^2 - 3\frac{\dot{H}}{\alpha_b} & = 0,\\
\frac{\ddot{\phi_b}}{\alpha_b^2}+\left(3H-\frac{\dot{\alpha_b}}{\alpha_b^2}\right)\frac{\dot{\phi_b}}{\alpha_b}+ \frac{dV}{d\phi}(\phi_b)  & = 0,
\end{aligned}\right .
    \label{eq:backgroundEq}
\end{equation}
where the Hubble rate is 
\begin{equation}
    H \equiv -\frac{1}{3} K_b = \frac{1}{\alpha_b}\frac{\dot{a}}{a}\,.
\end{equation} 
The background constraint equations are satisfied by definition when designing a model of inflation.

In this formalism, and before any gauge choice is made, the constraints \eqref{eq:cons_eq} linearise as 
\begin{equation}
\left \{
\begin{aligned}
        4\Delta \Phi-4H \kappa-2M_{Pl}^{-2}\delta \rho = 0, &\\
         -2\partial_i\left[\Delta \chi+\kappa+-\frac{3}{2}M_{Pl}^{-2}\frac{\dot{\phi_b}}{\alpha_b}\delta \phi \right]= 0, &\\
\end{aligned}\right .
    \label{eq:ConstraintsCPT}
\end{equation}
where the first order perturbation of the extrinsic curvature trace is
\begin{equation}
    \kappa  \equiv 3\left(\frac{\dot{\Phi}}{\alpha_b}+H \Psi\right)-\Delta \chi
\end{equation} 
with 
\begin{equation}
    \chi \equiv-\frac{a^2}{\alpha_b}(B-\dot{E})\,,
\end{equation} 
and the energy density perturbation is
\begin{equation}
   \delta\rho = \frac{\dot{\phi_b}}{\alpha_b}\frac{\delta\dot{\phi}}{\alpha_b}+\frac{dV}{d\phi}\bigg|_{b}\delta\phi-\frac{{\dot{\phi_b}}^2}{\alpha_b^2}\Psi \,.
\end{equation}

To make further progress, a gauge must be chosen and the constraints must be solved in that gauge to define physically admissible initial data for the dynamical variables. In certain gauges, the linearized constraints can be straightforwardly solved in terms of a linear gauge-invariant perturbation variable \cite{Launay24}. For instance, given the solution to the Mukhanov-Sasaski equation for the curvature perturbation on comoving hypersurfaces in Fourier space
\begin{equation}
\left \{
\begin{aligned}
\ddot{{\cal R}}_k + H(3-\varepsilon_2)\dot{{\cal R}}_k+ \frac{k^2}{a^2}{\cal R}_k = 0,\\
         {\cal R} = \Phi + \frac{H}{\dot{\phi_b}}\delta \phi,
    \end{aligned}\right .
    \label{eq:MS}
\end{equation}
where  the slow-roll parameters are $\varepsilon_{i+1} = - (\alpha_b H)^{-1}d_t \ln \varepsilon_i$ and $\varepsilon_0 = H$, one infers from \eqref{eq:ConstraintsCPT} that in the comoving gauge ($\delta\phi^{\rm co} = E^{\rm co}=  0$)
\begin{equation}
\left \{
\begin{aligned}
    \Phi^{\rm co}    & = {\cal R}, \\
    \Psi^{\rm co}  & = -H^{-1}\dot{{\cal R}},  \\
    B^{\rm co}  & = k^{-2}\varepsilon_1\dot{\cal R}  +(a^2H)^{-1}{\cal R}.  \\
\end{aligned}\right .
    \label{eq:ComGauge}
\end{equation}

However, the geodesic gauge (equivalent to synchronous gauge in a perturbative framework), or more generally a NR gauge choice which imposes $\Psi = \Psi_G$ and $B = B_G$, for arbitrary chosen functions $\Psi_G$ and $B_G$, is the easiest to evolve numerically. Therefore, despite not being the most widely used in cosmology in the past couple of decades, we opt for a geodesic gauge. Solving the constraints directly in this gauge is algebraically more involved than in the comoving gauge. An easier route is to use the comoving gauge relations eqns. \eqref{eq:MS} and \eqref{eq:ComGauge} as well as the gauge invariant Bardeen potentials. In the comoving gauge, those are (anisotropic stress is zero)
\begin{equation}
\left \{
\begin{aligned}
 \Phi_{B}  & =\Phi^{\rm co} + H\chi^{\rm co} =  -\varepsilon_1 H a^2 k^{-2}\dot{{\cal R}}, \\
         \Psi_{B}  & =  \Psi^{\rm co} - \dot{\chi}^{\rm co} = 
          -\varepsilon_1 H a^2 k^{-2}\dot{{\cal R}}\,.
    \end{aligned}\right .
    \label{eq:Bardeen}
\end{equation}
On the other hand, in a NR gauge 
\begin{equation}
\left \{
\begin{aligned}
 \Phi_{B}  & =\Phi^{\rm NR} + H\chi^{\rm NR}\,, \\
         \Psi_{B}  & =  \Psi_G - \dot{\chi}^{\rm NR} \,.
    \end{aligned}\right .
    \label{eq:BardeenSynch}
\end{equation}
 from which we can directly find, in such a gauge,
\begin{equation}
\left \{
\begin{aligned}
 \chi^{\rm NR}    & = \displaystyle\int^t_{t^{\circledast}} (\Psi_G  -\Psi_B  ) dt' + \chi_0^{\rm NR},\\
     E^{\rm NR}   & = \displaystyle\int^t_{t^{\circledast}}  [B_G  +a^{-2}\chi^{\rm NR}   ]dt' +E_0^{\rm NR},   \\
    \Phi^{\rm NR}    & = \Phi_{B}  -H \chi^{\rm NR},    \\
    \delta\phi^{\rm NR}    & = \sqrt{2\varepsilon_1 }M_{Pl}[{\cal R}   - \Phi^{\rm NR}  ]\,,\\
\end{aligned}\right .
    \label{eq:ADMgaugeDecomp}
\end{equation}
where $\Phi_B$ and $\Psi_B$ are given in terms of $\dot{\mathcal{R}}$, see eq.~\eqref{eq:Bardeen}. Therefore, all relevant quantities in this gauge have been expressed in terms of the single dynamical gauge-invariant scalar degree of freedom $\mathcal{R}$ and its time derivative $\dot{\mathcal{R}}$. The fields $\chi_0(\mathbf{x})$ and $E_0(\mathbf{x})$ are arbitrary initial space-dependent fields and their existence reflects the fact that the family of geodesic gauges is famously under-determined \cite{kodama84}. In this work, we set $\chi_0 = E_0 = 0$ at the start of the GR simulation $t^{\circledast}$, which will greatly simplify our initial conditions. 

As we will explicitly see, these relations and the associated time derivatives provide everything required to generate data for $(\gamma_{ij}^{\rm NR}, K_{ij}^{\rm NR}, \phi^{\rm NR}, \Pi^{\rm NR})$ as the sum of background inflation and these perturbations. Note that although $\Psi_G$ and $B_G$ are free to choose analytically or dynamically, they still need to be within perturbation theory initially to satisfy the constraints, which is why it is referred to as \textit{a small gauge} choice. The numerical evolution in this work assumes such a gauge choice.
\begin{figure*}
    \centering
\includegraphics[width=0.9\linewidth]{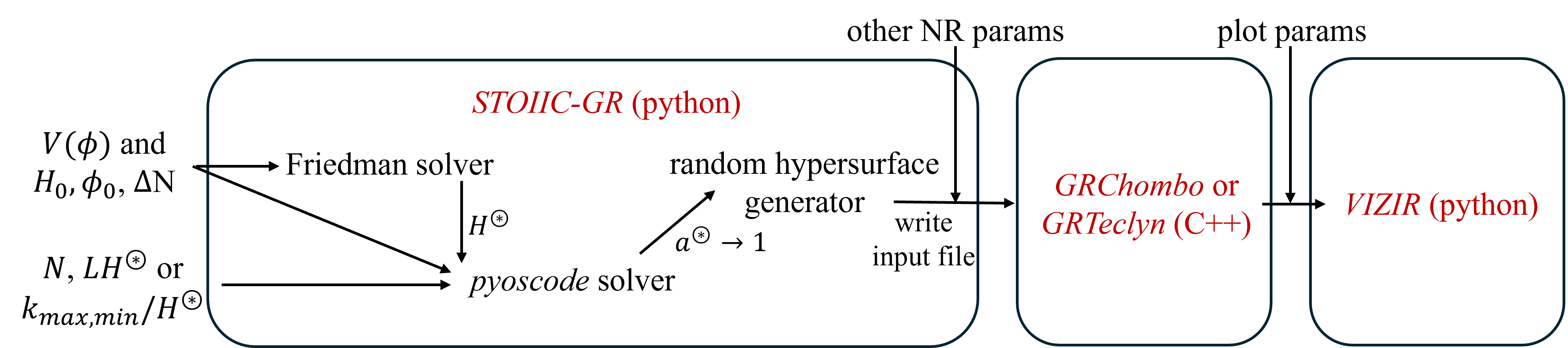}
    \caption{\justifying Full numerical pipeline. The user specifies in chosen units a potential $V$ either entirely or as a functional of parameters, a Hubble constant $H_0$ to start with or an initial field value $\phi_0$ (see \ref{subsec:background}), a number of background efolds $\Delta \mathcal{N}$ between the Bunch-Davies vacuum initial conditions and the starting time of the GR simulation, $N$ the number of pixels per dimension to generate and get simulated, the length $LH^{\circledast}$ or the Fourier range $k_{max,min}/H^{\circledast}$ of the simulated box in terms of the Hubble radius at $t^{\circledast}$. The user can also specify parameters related to NR evolution (stepping, damping, adaptive mesh refinement, etc), or plotting requests from the final extractions. }
    \label{fig:pipeline}
\end{figure*}
\subsection{Numerical Relativity dictionary for cosmology}
To solve partial differential equations, it is necessary to reformulate them into a well-posed system with first-order-in-time partial differential equations, ensuring a unique solution. BSSN equations are now the common GR equations for these purposes \cite{BSSN_SN,BSSN_BS}. These are written explicitly in Appendix \ref{app:BSSN}.

In this formalism, the metric is written as in eq. \eqref{eq:ADMmetric} up to a minus sign convention for the shift $\beta^i$. The spatial metric is decomposed into a conformal factor $X$ and a conformal metric $\tilde{\gamma}_{i j}$
\begin{equation}
\left\{
\begin{aligned}
    \gamma_{i j} & =\frac{1}{X} \tilde{\gamma}_{i j}, \\ X & =\left(\operatorname{det} \gamma_{i j}\right)^{-\frac{1}{3}},
\end{aligned}\right .
    \label{eq:ConformalFactor}
\end{equation}
so that the conformal metric has unit determinant $\operatorname{det} \tilde{\gamma}_{i j}=1$. 
The extrinsic curvature is decomposed into its trace $K=$ $\gamma^{i j} K_{i j}$ and its conformally rescaled traceless part ($\tilde{\gamma}^{i j}\tilde{A}_{i j}=0$ and $X\tilde{K}_{ij} = \tilde{A}_{ij}$) as
\begin{equation}
K_{i j}=\frac{1}{X}\left(\tilde{A}_{i j}+\frac{1}{3} K \tilde{\gamma}_{i j}\right).
    \label{eq:ConformalK}
\end{equation}
Finally, an intermediary quantity is defined: the conformal contracted connections defined as $\tilde{\Gamma}^i=\tilde{\gamma}^{j k} \tilde{\Gamma}_{j k}^i$ where $\tilde{\Gamma}_{j k}^i$ are the Christoffel symbols of the conformal metric $\tilde{\gamma}_{i j}$. 

This means that the initial hypersurface needs to specify the dynamical quantities  $X$, $\tilde{\gamma}_{ij}$, $K$, $\tilde{A}_{ij}$, and $\phi$, $\Pi$.\footnote{$\tilde{\Gamma}^i$ being an intermediary quantity, function of the others.} Previous eq. \eqref{eq:ADMgaugeDecomp} can thus be translated into its (linear) BSSN version (see Appendix \ref{app:dicBSSN} for conversions) in a NR gauge as 
\begin{equation}
\left\{\begin{aligned}
 \delta{\Pi}^{\rm NR} &  = \delta\dot{ \phi}^{\rm NR}-\dot{{ \phi}_b}\Psi_G,\\
      \delta X^{\rm NR} & = a^{-2} [2\Phi^{\rm NR} -\frac{2}{3}\nabla^2 E^{\rm NR}], \\
    \delta\tilde{\gamma}_{ij}^{\rm NR} & =  2(\partial_i\partial_j-\frac{1}{3}\delta_{ij}\nabla^2)E^{\rm NR},\\ 
    \delta K^{\rm NR} &  =  3(\dot{\Phi}^{\rm NR}+H\Psi^{\rm NR})-\Delta \chi^{\rm NR},\\
    \delta\tilde{A}_{ij}^{\rm NR} &   = (\partial_i\partial_j-\frac{1}{3}\delta_{ij}\nabla^2)(B_G-\dot{E}^{\rm NR}).\\ 
\end{aligned}\right .
\label{eq:eqDictionary}
\end{equation}
Substituting the previous perturbations eq. \eqref{eq:ADMgaugeDecomp} and evaluating at $t^{\circledast}$ yields
\begin{equation}
\left\{\begin{aligned}
    \delta\phi^{\rm NR,\circledast}_{\vec{k}} & = \sqrt{2\varepsilon_1^{\circledast}}M_{Pl}({\cal R}_k^{\circledast}  - \Phi_{Bk}^{\circledast}), \\
    \delta\Pi^{\rm NR,\circledast}_{\vec{k}} & = -\sqrt{2\varepsilon_1^{\circledast}}H^{\circledast}M_{Pl}\Psi_G^{\circledast} +\sqrt{2\varepsilon_1^{\circledast}}M_{Pl}(\dot{{\cal R}}_k^{\circledast} - \dot{\Phi}_{Bk}^{\circledast})\\
    & -\frac{1}{\sqrt{2}}\varepsilon_2^{\circledast}\sqrt{\varepsilon_1^{\circledast}}H^{\circledast}M_{Pl}({\cal R}_k^{\circledast}  - \Phi_{Bk}^{\circledast}),\\
     \delta X^{\rm NR,\circledast}_{\vec{k}} & = 2\Phi_{Bk}^{\circledast} ,\\
    (\delta\tilde{\gamma}_{ij})^{\rm NR,\circledast}_{\vec{k}} &  = 0,\\ 
    \delta K^{\rm NR,\circledast}_{\vec{k}} & = 3(\dot{\Phi}_{Bk}^{\circledast}+H^{\circledast}\Psi_{Bk}^{\circledast}),\\
    (\delta\tilde{A}_{ij})^{\rm NR,\circledast}_{\vec{k}} & =  0,\\ 
\end{aligned}\right .
\label{eq:eqInitialConditions}
\end{equation}
where the Bardeen potentials are expressed in terms of $\dot{\mathcal{R}}$ via eq.~\eqref{eq:Bardeen}. In conjunction with that, the backgrounds are set as $X_b^{\circledast} = a^{-2}(t^{\circledast}) = 1$, $(\tilde{\gamma}_{b})_{ij}^{\circledast} = \delta_{ij}$, $K_b^{\circledast}=-3H^{\circledast}$ and $(\tilde{A}_b)_{ij}^{\circledast} = 0$. We just proved that initial conformal flatness ($\tilde{\gamma} = \delta$) is valid in this gauge at first order. However, looking at eq. \eqref{eq:eqDictionary} suggests that $\delta X$ cannot be $0$ initially simultaneously and vice-versa.

In summary, we have solved the linear constraints of GR by expressing all quantities on the initial hypersurface in terms of a variable very well known to cosmologists, the linear, gauge-invariant curvature perturbation $\mathcal{R}$ {and its time derivative $\dot{\mathcal{R}}$} - see equation \eqref{eq:eqInitialConditions}.

\subsection{Classical simulation \label{subsec:clasical}}
 In most inflationary scenarios, it is assumed that scalar perturbations are sourced by quantum initial conditions coming from the vacuum energy on small scales.
Quantization is usually written down explicitly as an expansion in the Fourier modes of an operator such as
 \begin{equation}
  \hat{\cal O} = {\cal F}^{-1}\{ {\cal O}_{\vec{k}}\hat{a}_{\vec{k}}\} + h.c,
         \label{eq:ModesExpGeneral}
 \end{equation}
 using each mode's harmonic annihilation and creation operators such that $[\hat{a}^{}_{\vec{k}},\hat{a}_{\vec{k}'}^\dagger ] =  \delta^{(3)}(\vec{k}-\vec{k}')$ and any other commutator being $0$. The operators, the Fourier amplitudes of which are written in eq. \eqref{eq:eqInitialConditions} or that of ${\cal R}$, are constructed this way. 
 
 However, we do not (and cannot) input a quantum operator to a classical simulation. 
Fortunately, it is common to assume that the operators become random variables (by dropping the non-commutating parts) above a certain scale called the comoving Hubble radius: $k \leq aH$ \cite{Starobinsky_stochastic_1988}. More recently \cite{Launay2024bis}, it was also argued that inputting slightly subHubble scales could also be used for examining the signatures of (cubic) interactions. 
In that case, eq. \eqref{eq:ModesExpGeneral} can be summarised by its spacetime correlation functions from which random realisations can be drawn. The statistics of such a harmonic mix are Gaussian and thus are completely described by the two-point functions. This is equivalent to writing an expansion of stochastic modes for each operator $i$
\begin{equation}
  \boldsymbol{{\cal O}}_i = {\cal F}^{-1}\{ {\cal O}_{i,\vec{k}}\boldsymbol{\alpha}_{\vec{k}}\} + h.c,
         \label{eq:StoModesExpGeneral}
\end{equation}
where $\boldsymbol{\alpha_{\vec{k}}}$ are random variables such that
\begin{equation}
           \langle \boldsymbol{\alpha_{\vec{k}}}\boldsymbol{\alpha_{\vec{k}'}}^* \rangle_{\mathbb{P}}  = C \delta^{(3)}(\vec{k}-\vec{k}'),
\label{eq:stocomm}
\end{equation}
where $C$ is a normalisation constant (see section \ref{sec:randomdraw} and appendix \ref{app:random} for determination). It is important to remember that there is only one dynamical scalar degree of freedom, as seen from the Mukhanov-Sasaki equation \eqref{eq:MS}. This is why {all quantities on the initial} hypersurface can be fully expressed in terms of ${\cal R}_k$ {and $\dot{\cal{R}}_k$ as} in eq.~\eqref{eq:eqInitialConditions} in Fourier amplitudes space, but also in operator space in terms of $\boldsymbol{{\cal R}}$: there is only one set of $\{\boldsymbol{\alpha}_{\vec{k}}\}$ for all operators.

This analytical material is essentially all that is needed to start a full GR simulation and learn about full non-linearities and non-perturbative features during inflation.
Note that completely equivalent reasoning can be made on graviton perturbations (tensor modes) sourced by the vacuum and has been used in parallel work \cite{Florio2024}, although it is two additional degrees of freedom and the constraints are trivially satisfied at first order, which does not necessitate the reasoning of the present paper. 

\section{Numerical methods \label{sec:nummethods}}
In this section, we detail our numerical methods and pipeline, represented in Figure \ref{fig:pipeline}. Here is a summary for the busy reader who wants to skip the technical aspects of our numerical work:
\begin{itemize}
\item[\ding{118}] Friedmann and Mukhanov-Sasaki equations are evolved from the imposition of the Bunch-Davies vacuum to the start of the GR simulation.
\item[\ding{118}] A single 3D random box is drawn and given the right spectrum for each of the BSSN quantities of eqn \eqref{eq:eqInitialConditions}. The constraints are accurately satisfied at first order in perturbation theory (violation only at second order and beyond). The included wavenumbers go from slightly subHubble to a few times the Hubble radius.
\item[\ding{118}] This stochastic box is then evolved using a full GR simulation. The main software used is \textit{GRChombo} \cite{Andrade2021} but the results have been successfully recovered using the GPU-enabled ported software \textit{GRTeclyn} \cite{GRTeclyn}.
\item[\ding{118}] Quantities of interest are extracted after a few efolds.
\end{itemize}
\subsection{Designing an inflationary patch (STOIIC-GR) \label{subsec:designpatch}}
The parameter space for a simulation with fixed evolution mostly boils down to dimensionless physical initial conditions and resolutions in space and time. However, these choices can influence or constrain each other, especially when studying inhomogeneous inflation. 

Our goal is to observe the inflaton's self-interaction in GR. In a perturbative framework, which is what we will always start from, these interactions are key for the scales crossing the Hubble radius defined previously because the horizon crossing entails a spectral turnover from a subHubble to a superHubble slope. This means that our simulation needs to showcase at least one mode crossing this horizon: the box resolution and the input analytical spectrum need to be such that $k_{max}^{box},k_{max}^{\cal R}\geq(aH)^{\circledast}$ but also such that gradients can resolve the perturbations $k_{max}^{box}> k_{max}^{\cal R}$. Written that way, wavenumbers $k$ will always be taken in comoving Fourier space and so will our spatial coordinates be in the GR simulation, those of eq. \eqref{eq:SVTmetric}.  Our simulations will thus always see the Hubble radius $R = (aH)^{-1}$ shrinking in the box until it disappears beyond the discrete resolution.

For a fixed $k_{max}^{\cal R}$ and the gradients-friendly choice of $k_{max}^{box}$, the computational resources will determine the maximal size of the box (or $k_{min}$) through the maximum number $N$ of pixels per dimension we can run the code with. Beyond that size, we assume that the simulated universe is made of periodic patches such as this one. This is equivalent to giving a null spectrum to that range of superscales or a known spectrum de-correlated from those smaller scales.  This is a very good approximation in the case of inflation if the patch is of size a few times $R^{\circledast}$ because of the separate universe behaviour, where separate patches barely influence each other and can be looked at as different random draws of the same stochastic box of finite size \cite{salopek_nonlinear_1990, salopek_stochastic_1991}. 

If $N$ is big enough to have both a superHubble-sized box and subHubble scales, it is tempting to take $k_{max}^{box},k_{max}^{\cal R}\gg (aH)^{\circledast}$ similarly to what was achieved in the Friedmann lattice code of \cite{Caravano2024}. From a numerical point of view, nothing hampers the treatment of that regime. From a physical point of view, we {claim} that it is not possible to go too deep sub-Hubble without encountering well-known UV divergences of the Mukhanov solutions. We believe that the choice of gauge can also change how sub-Hubble one can go. The key difference with \cite{Caravano2024} is that of the full GR treatment versus the Friedmann local one, which {in practice translates to using} the Mukhanov-Sasaki linear equation of the Fourier modes. It is fine to leave perturbation theory (and these lattice Friedman approaches certainly do so) and use the Mukhanov-Sasaki evolution to recover the right late-time perturbations, as long as no back-reaction is investigated. In our case, going too deep would probably drive the evolution equations away from the background and linear solutions when using the full GR action and equations, but most importantly perturbation theory breaks down and so does our initialisation - see Appendix \ref{app:UVdep}. This problem is fundamental and links to the debate on UV renormalisation (see \cite{pla23} for a recent treatment); the vacuum energy might need to be removed as only its late products are gravity-friendly. 
Until this topic is completely addressed with a {satisfactory treatment of the gravitational effects of deep UV modes}, this work will simply be cautious by applying a simple momentum cutoff (see $\sigma_{max}$ below), {to stay within perturbation theory and negligible backreaction at the initial time}.

We summarise all our recommendations as
\begin{equation}
\left\{
    \begin{aligned}
         k_{max}^{\cal R} & = \sigma  /R^{\circledast}, \\
         dx & \leq \frac{2\pi R^{\circledast} }{n\sigma},  \\
         1 \leq \sigma& \leq \sigma_{max},
    \end{aligned}\right .
    \label{eq:designbox}
\end{equation}
where $n$ is the number of pixels to cover the associated smallest wavelength ($n\geq 4$ in our work, using Runge-Kutta 4 and its stencils) and ensure that initial data is smooth enough. $\sigma_{max}$ is the scenario-dependent perturbativity bound above which we cannot guarantee that our simulation will handle {the corresponding modes perturbatively}. We also assumed that the IR sector is such that $k_{min}^{\cal R} = k_{min}^{box} = \frac{2\pi}{Ndx}$. This setup also automatically imposes at least a Hubble-sized initial box as it only takes $N\geq  \frac{2\pi}{dx}R^{\circledast}\geq 1$ to have $k_{min}\leq 1/R^{\circledast}$.

\subsection{Perturbative dynamics}
\subsubsection{Background solutions \label{subsec:background}}
The first choice to make is that of the background quantities of inflation and in particular of the background initial conditions. There is not always a unique way to impose certain constraints on the background dynamics. In this work and unless stated differently, we usually choose an initial $H_0$ and a potential $V_m(\phi)$ parametrised by an unknown parameter $m$ first, then we determine $\phi_0,m$ to match certain values of a slow-roll attractor. For instance, in a $V(\phi)=\lambda \phi^p $ inflation ($p\geq 0$ even), one can use the Friedmann equations to set
\begin{equation}
\left\{
\begin{aligned}
    \phi_0 & = -\left|\frac{ (3 - \varepsilon_1)pM_{Pl}}{ \sqrt{2  \varepsilon_1}  (\varepsilon_2 / 2 + \varepsilon_1 - 3)}\right|, \\
           \lambda & =  \left |\frac{\sqrt{2 \varepsilon_1}  (\varepsilon_2/ 2 + \varepsilon_1 - 3) }{ (3 - \varepsilon_1) pM_{Pl}}\right|^p M_{Pl}^2  H_0^2 (3 - \varepsilon_1),
\end{aligned}\right .
\label{eq:potentialposition}
\end{equation}
for a given set of slow-roll parameters $\epsilon_{1,2}$, notably $\varepsilon_2 = -2\varepsilon_1$ for the attractor. 

Another method we employed was to fix entirely $V(\phi)$ and pick $\phi_0$ at a point of interest (for instance before a feature). The field's velocity would then be fixed using the slow-roll limit expression $\dot{\phi_0}=-\frac{V_{,\phi}}{V}\displaystyle |_{\phi_0}M_{Pl}^2$, thus fixing $H_0$ through the Friedmann constraint, the last quantity required for solving. Both methods guarantee a slow-roll evolution.

We will refer to these two methods as  \textit{potential-position-fixing} and \textit{$H_0$-fixing} respectively.  The former though might be the most appropriate to relate to observational constraints, usually made on the shape of the potential, the efolding duration, and the spectral index.
Note that we do not claim that any of our models are appropriately scaled to match e.g. Planck data \cite{PlanckInflation}. The potentials of section \ref{sec:Applications} though should still have portions consistent with data.

As nicely expressed in \cite{Agocs20}, our pipeline evolves then the Friedmann equations \eqref{eq:backgroundEq} re-written as

\begin{equation}
\left\{
    \begin{aligned}
        \frac{d\ln{\Omega_k}}{d{\cal N}} & = 4 -2a^2 \Omega_kV(\phi),\\
        \Big(\frac{d\phi}{d{\cal N}}\Big)^2&  = 6 -2a^2\Omega_kV(\phi),
    \end{aligned}\right .
    \label{eq:rewriteFriedman}
\end{equation} 
where $\Omega_k = (aH)^{-2}= R^2$ is the curvature density in a flat universe. This is evolved for an arbitrary number of efolds $\Delta{\cal N}$ from the ICs described above. 

\begin{figure}
\includegraphics[width=0.95\linewidth]{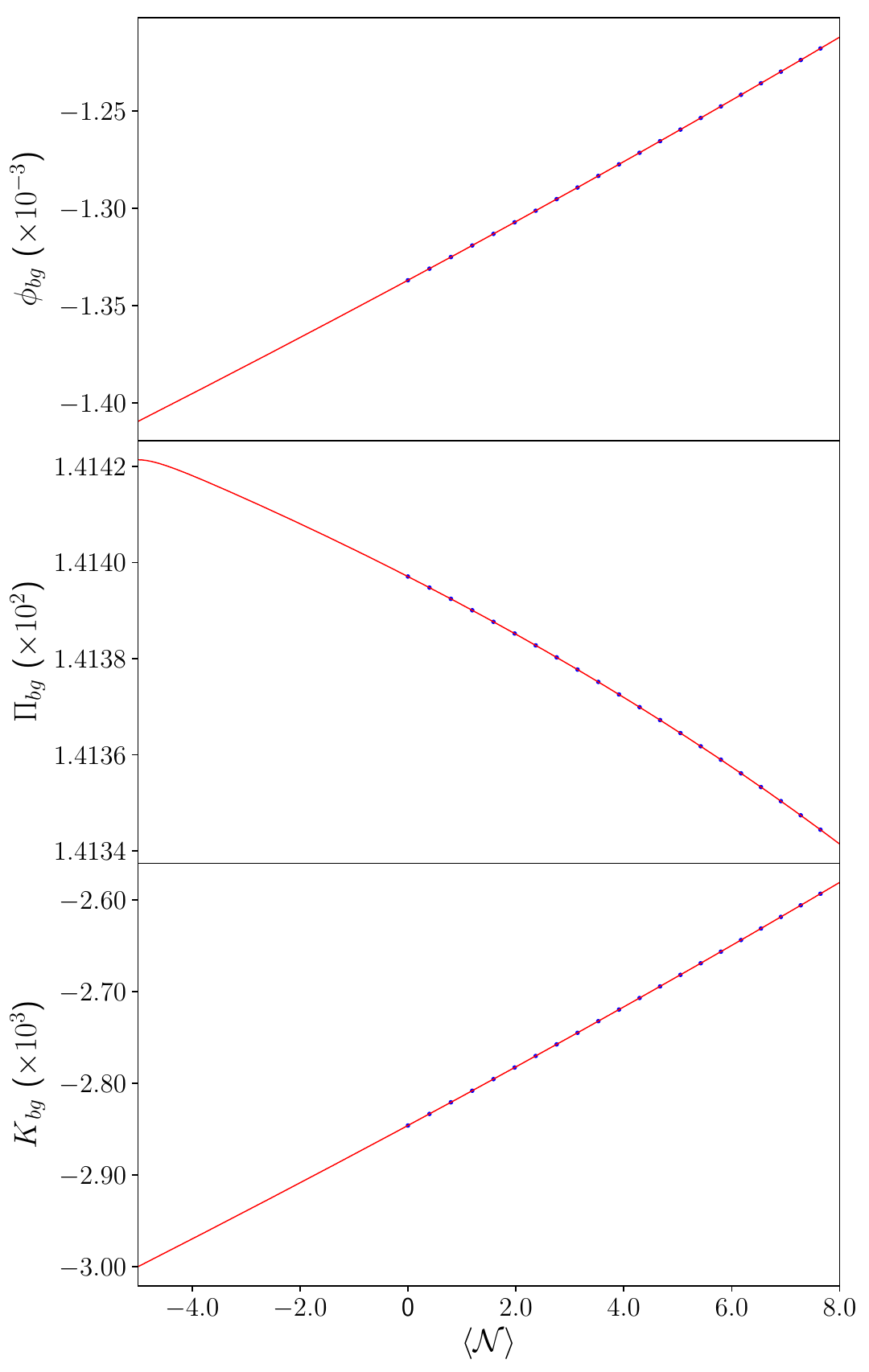}
\caption{\justifying Comparison of the Friedmann solution (solid red) and the mean inflaton (top), conjugate momentum (middle), and extrinsic curvature (bottom) fields extracted from the GR simulation (blue dots)  for the quadratic inflation case. $M_{Pl}=100$ units.}
\label{fig:meanquad}
\end{figure}
\begin{figure*}
    \centering
        \begin{subfigure}{0.32\textwidth}
        \centering
\includegraphics[width=\linewidth]{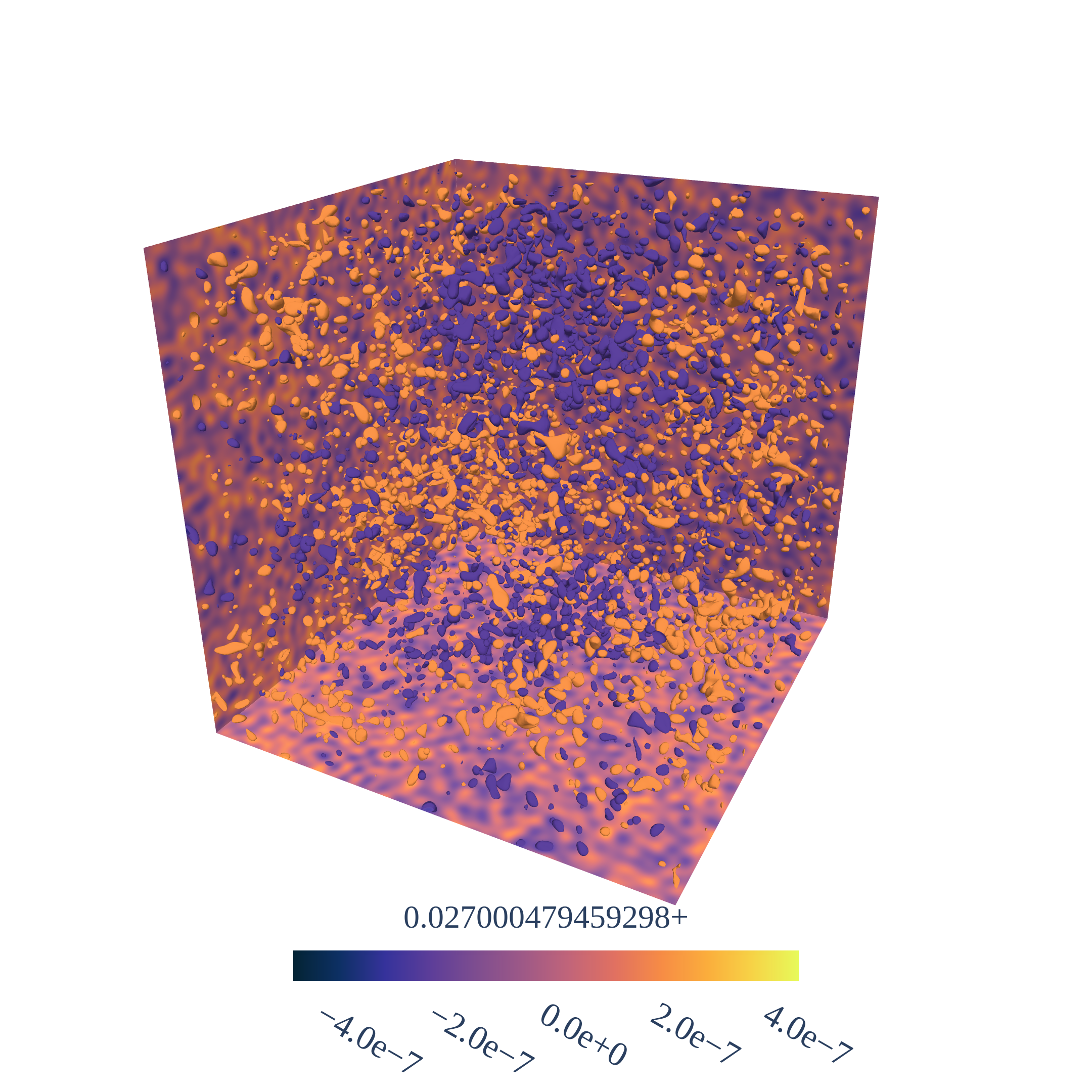}
    \caption{$\langle {\cal N}\rangle=0$}
    \label{fig:3Da}
    \end{subfigure}
    \hfill
\begin{subfigure}{0.32\textwidth}
        \centering        \includegraphics[width=\linewidth]{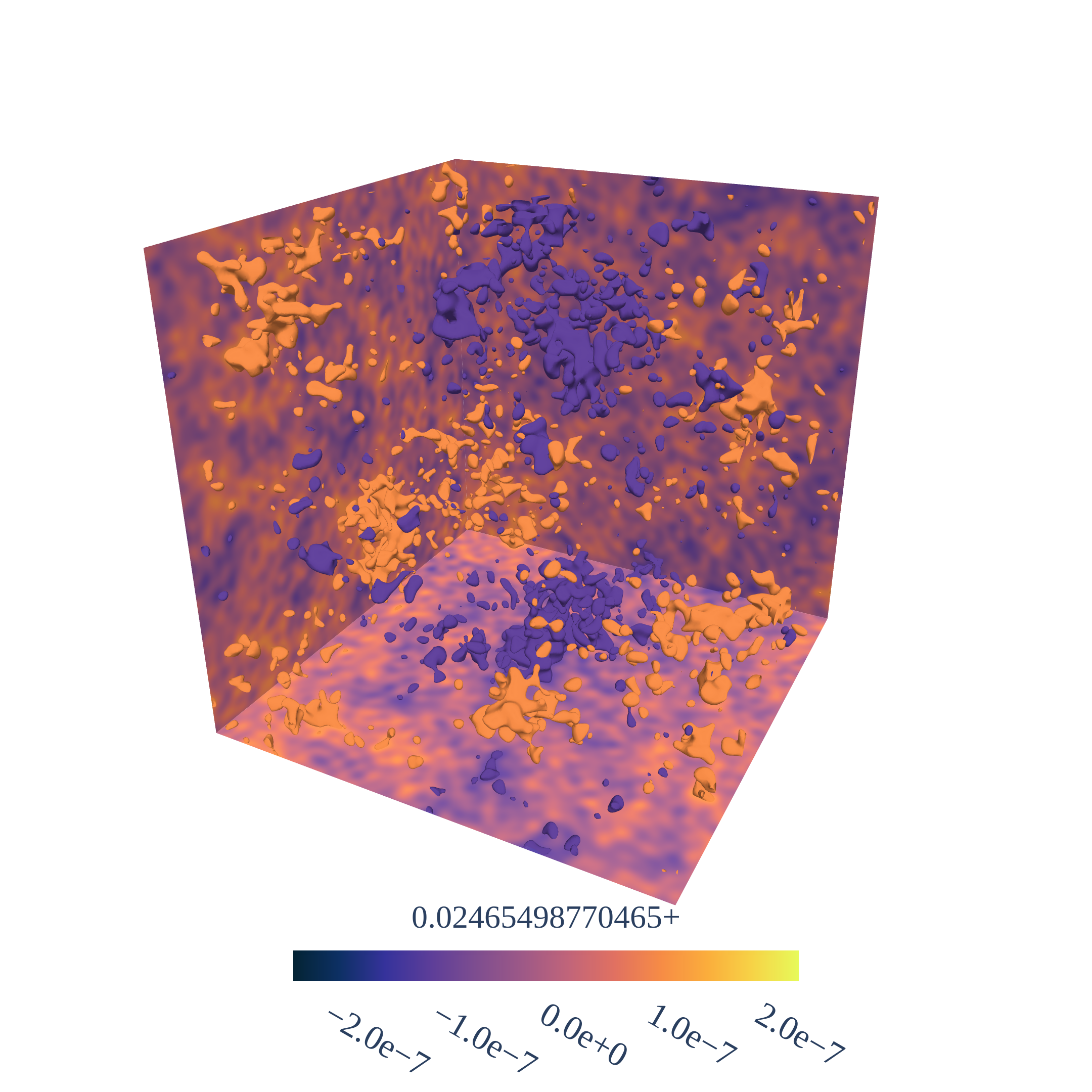}
        \caption{$\langle {\cal N}\rangle=3.91$}
        \label{fig:3Db}
    \end{subfigure}
\hfill
\begin{subfigure}{0.32\textwidth}
        \centering
        \includegraphics[width=\linewidth]{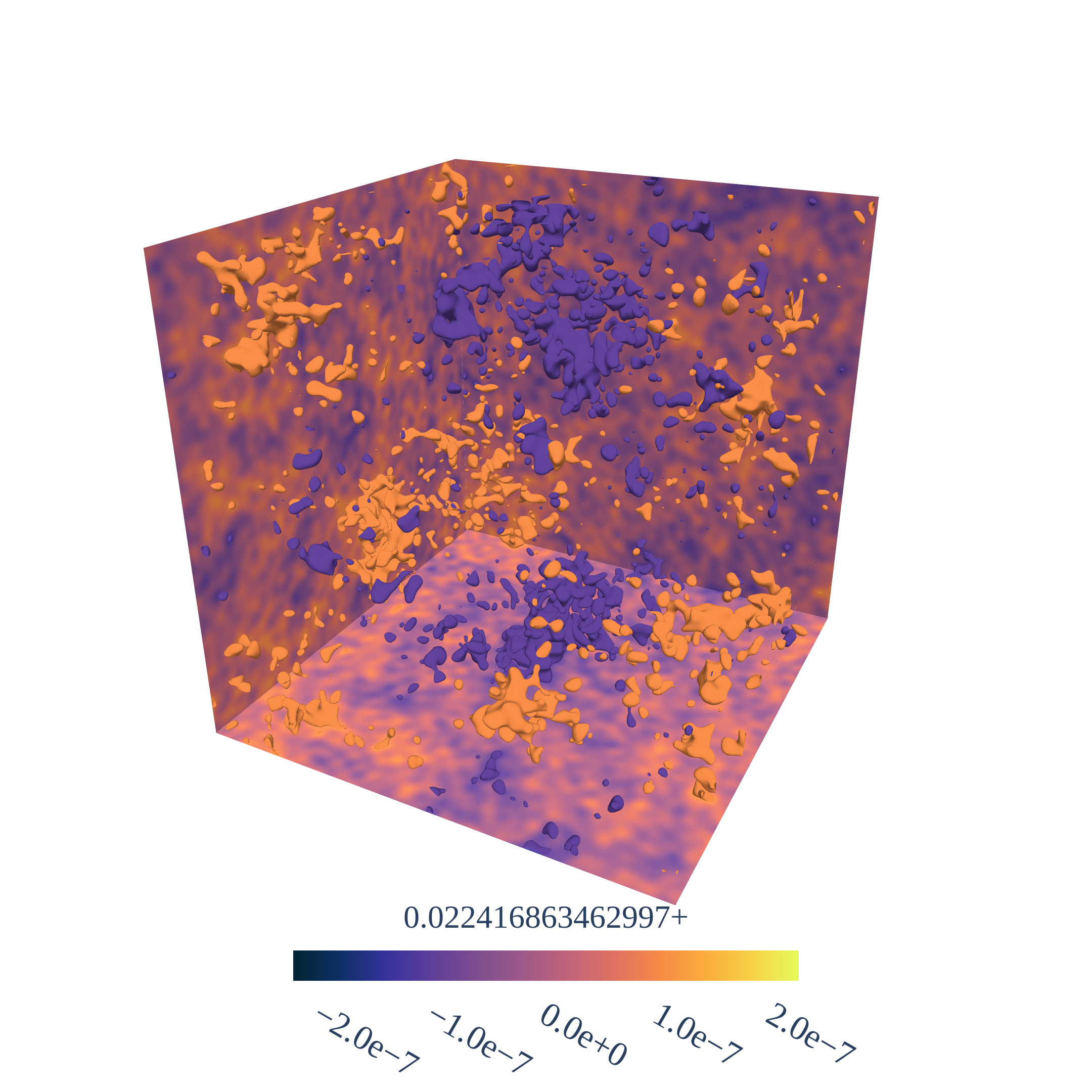}
        \caption{$\langle {\cal N}\rangle=7.64$}
        \label{fig:3Dc}
    \end{subfigure}
\caption{\justifying  Energy densities evolution (left to right) of the simulated patches for quadratic inflation. 2D boundary heat maps and volumic isosurfaces share the same color bar, but not across plots. Isosurfaces are at $\pm 2.5$std. $M_{Pl}=100$ units. }
\label{fig:3Dplots0}
\end{figure*}
\subsubsection{Linear solutions (\textit{pyoscode})}
Once the background dynamics have been solved, it is possible to solve eq. \eqref{eq:MS}, the solution of which might be highly oscillatory at high wavenumbers. This can be numerically problematic and requires an adequate solver such as \textit{pyoscode} introduced in \cite{Agocs20,agocs20bis}. The code was primarily developed for the Mukhanov-Sasaki equation and this part of our pipeline adapts some of the example codes to our needs.

The evolution of eq. \eqref{eq:rewriteFriedman} for  $\Delta{\cal N}$ efolds lands on realistic values of the Hubble parameter, scale factor, field, and its momentum at $t^{\circledast}$, the GR simulation starting time:  $(H^{\circledast}, a^{\circledast}, \phi^{\circledast},\dot{\phi}^{\circledast})$. From that, our choice for the range of modes with respect to the Hubble radius at that time can be made explicit. In particular,  if following the previous recommendations of eq. \eqref{eq:designbox} and if $\Delta{\cal N}$ is big enough, these should all be deep subHubble back at $t_0$ when the Bunch-Davies vacuum will be imposed.  This is why we recommend $\Delta{\cal N}\geq 2$.

The discrete set of $\{k_{min}^{\cal R}(R^{\circledast}),...,k_{max}^{\cal R}(R^{\circledast})\} $ gives as many Mukhanov-Sasaki equations to solve using this code. After that, we have all the values needed for ${\cal R}_k$ to impose the associated spectrum to our grid.

\subsubsection{Random hypersurface \label{sec:randomdraw}}

One way to impose a spectrum ${\cal P}_k = |f_k|^2$ ($f_k$ being the Fourier amplitude) to a Gaussian and independent random 3D draw in real space is to multiply the Fourier amplitude of the latter by $\sqrt{{\cal P}_k} = |f_k|$. Equivalently in the large $N$ limit, the random draw part can be done directly in Fourier space using complex normal draws $\mathbb{C}{\cal N}(0,1)$ and imposing hernitian symmetry. In this work, the former method is adopted. 

This work imposes several Fourier amplitudes $f_k$, defined in eq. \eqref{eq:eqInitialConditions}, all functions of ${\cal R}_k$ and its time derivative. To reach the perfect satisfaction of the linear constraints \eqref{eq:ConstraintsCPT}, an adaptation of the usual method is needed because the sign of perturbations for every quantity matters against others. Hence, our amplitudes will be $f_k$ as opposed to $|f_k|$, which does not impact the validity of the method. In this sense, we are literally applying formula \eqref{eq:StoModesExpGeneral}. It is also worth recalling from section \ref{subsec:clasical} that there is only one 3D random draw and degree of freedom despite having multiple stochastic quantities and so all these quantities are cross-correlated accordingly. 

When working with discrete Fourier transforms, power spectra, and different conventions, reaching the right normalisations and units-invariance can quickly become challenging. Appendix \ref{app:random} makes transparent the numerical construction of our perturbations using the method described above and the link with QFT. In particular, one should remember that cosmological power spectra from QFT have surprising units because of the ones implicitly carried by annihilation and creation operators. Our generator of perturbation has been tested successfully with respect to the linear and full GR constraints and showed on multiple occasions that switching a sign or any slight modification of the constraint formulae would show a violation of these (see constraints plot in applications). The satisfaction of the constraints is the result of a highly precise and cautious method.

The transition to the GR simulation is the next step, once constraints and spectra have been checked. This is usually done by producing an input parameters file for the GR simulation, together with a grid file it can start from. Before switching to NR, our pipeline also performs a rescaling of the scale factor, the Planck mass, and their dependencies to reach a range of numbers within double precision, valid for the entire duration of the simulation. Typical choices include $a^{\circledast}=1$ and $M_{Pl}=100$ (equivalent to having quantities in $0.01M_{Pl}$ units of masses).  The latter choice will reveal as particularly useful to get second-order perturbative violations of the constraints above the numerical accuracy (see applications in section \ref{sec:Applications}).

\subsection{Numerical relativity \label{NR}}

\begin{figure}
    \centering
\includegraphics[width=\linewidth]{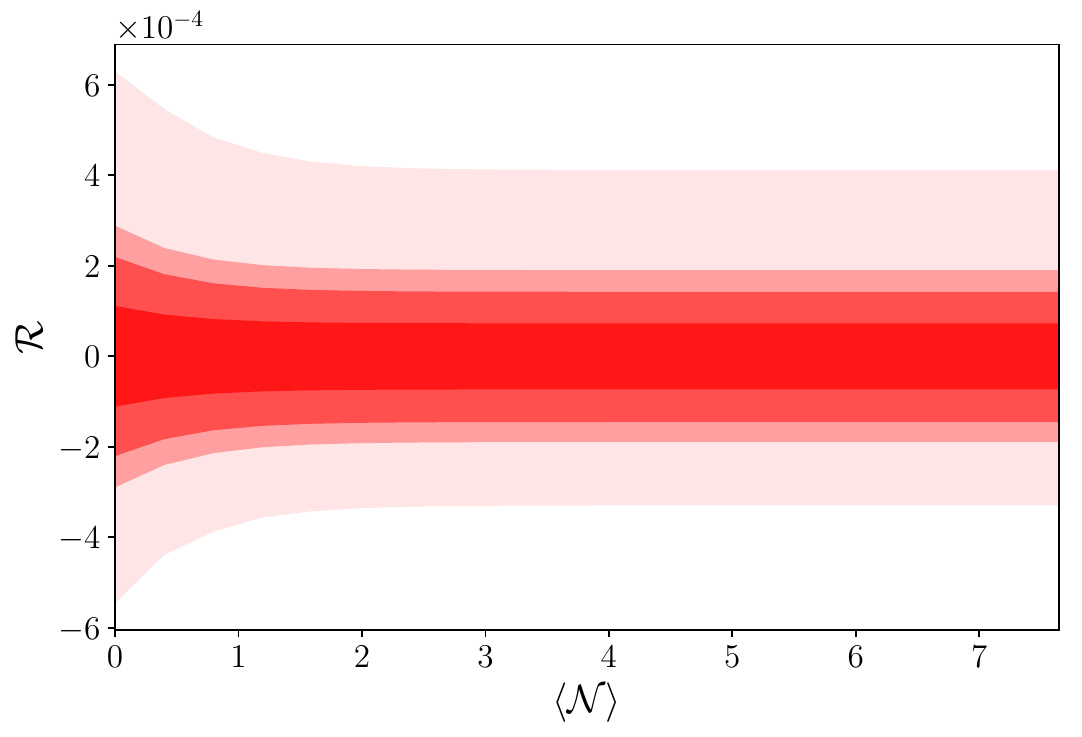}
    \caption{\justifying Contrast of ${\cal R}$ during quadratic inflation using $68$, $95$, $99$ and $100$ percentile contours (dense red to transparent red). $M_{Pl}=100$ units.}
    \label{fig:quad_rlin}
\end{figure}
\begin{figure*}
    \centering
\includegraphics[width=0.7\linewidth]{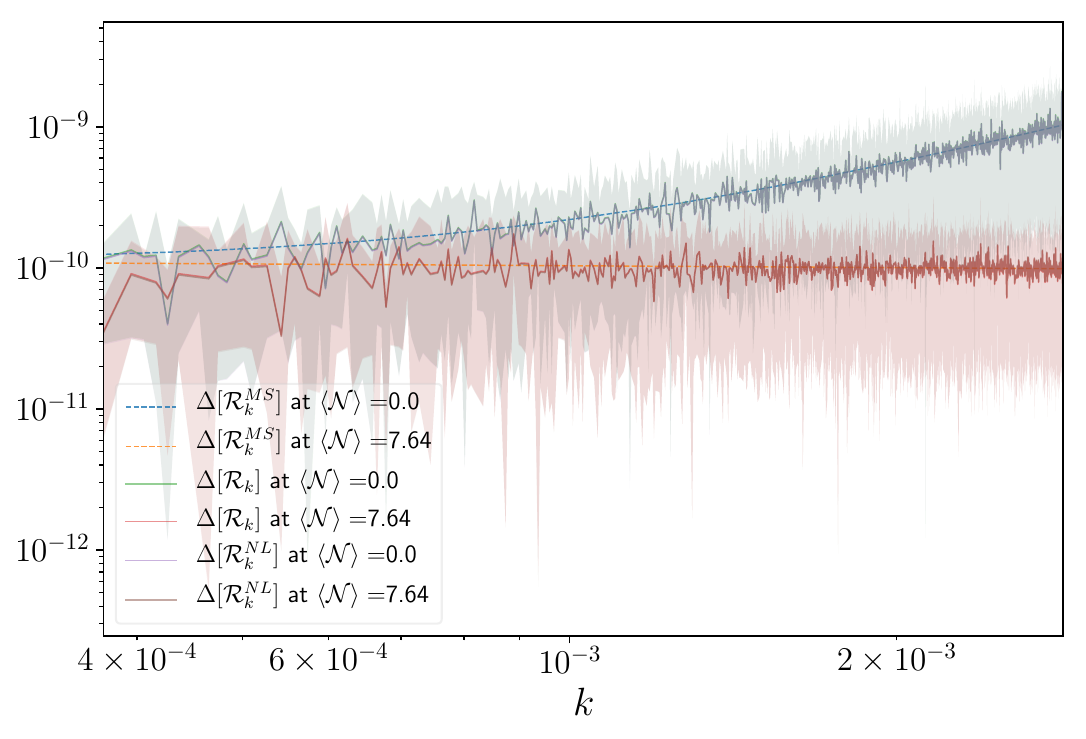}
    \caption{\justifying Linear (dashed, ${\cal R}^{MS}$) and extracted (solid, ${\cal R}$) binned dimensionless spectra ($\Delta[{\cal R}]= \frac{k^3}{2\pi^2} |{\cal R}_k|^2$) compared to extracted ${\cal R}^{NL}$, scalar built from the fully non-linear generalisation ${\cal R}_i$ in the quadratic inflation case{: perfect superimposition at initial (upper half) and final (lower half) times ({\textcolor[HTML]{95d095}{\rule{1em}{0.5em}}} $+$ {\textcolor[HTML]{cab3de}{\rule{1em}{0.5em}}} $=$ \textcolor[HTML]{608475}{\rule{1em}{0.5em}}, {\textcolor[HTML]{ea9393}{\rule{1em}{0.5em}}} $+$ {\textcolor[HTML]{c6aba5}{\rule{1em}{0.5em}}} $=$ \textcolor[HTML]{b13e3a}{\rule{1em}{0.5em}})}. Contours account for the $1\sigma$-range of points in each bin. Displayed Fourier bins are those of the simulation mesh which have more than 10 realisations and which are smaller than the spectral cutoff of the input ($k_{max}^{\cal R} = 3\times 10^{-3}$). $M_{Pl}=100$ units.}
\label{fig:specquad}
\end{figure*}

GRChombo and more recently GRTeclyn are two open-source codes from the GRTL collaboration \cite{GRTL} for evolving the BSSN and CCZ4 equations of numerical relativity \cite{Andrade2021, radia_lessons_2022}. Both were built to make use of hybrid MPI/OpenMP parallelism and vector intrinsics to achieve the best performances on the latest HPC architectures while offering adaptive mesh refinement (AMR) technology. The latest software was motivated to enable the use of GPUs thanks to the AMReX library \cite{amrex} and has thus allowed consequent speed-ups. In this work, the pipeline can use either of them and lead to the same results. Though, as more established so far and as GRTeclyn is still under development, GRChombo is behind every simulation plot of this article.

Evolving full non-linear inhomogeneous inflation seems challenging in cosmology, but it is a rather light problem for NR (up to complications before and after evolving mentioned in \cite{aurrekoetxea24}). From this full NR library, we will not make use of the AMR or the CCZ4 damping to eliminate gauge artefacts \cite{CCZ41, CCZ42} but simply the BSSN equations \cite{BSSN_BS,BSSN_SN} (see Appendix \ref{app:BSSN}) with our initial conditions and periodic boundary conditions.

In BSSN codes, the numerical gauge is usually a 1+log-slicing evolving together with Gamma drivers \cite{baumgarte_shapiro_2010,radia_lessons_2022}, but our work only uses the sub-case of a geodesic observer, which,  in the cosmological language, is the synchronous gauge presented earlier in section \ref{subsec:Inhom-Constr}, with $\Psi_G = B_G = 0$. We will see that this gauge comes with some caveats.  

Regarding the time resolution in these coordinates, this work sets the Courant factor value as $dt/dx = 0.025 $ while only extracting $15$ to $50$ checkpoints. The typical size of our box is $256^3$. This size usually requires less than a few hours to run on $64$ to $152$ processors and still provides insightful results\footnote{Note that tests have also been successfully run with $1024^3$ in GRTeclyn but are computationally and memory-wise more demanding to run and post-process (see in-situ visualisation solutions of the GRTL collaboration, and future work).}. Convergence tests are available in subsection \ref{sec:correct}.

\subsection{The final values problem}
The previous software provides three-dimensional raw fields at multiple time slices and require efficient and careful extraction. We made the choice of using the \textit{python} \textit{YT} library \cite{YT}. Plotting was made flexible to some extent through our project {VIZ}ualizer for {I}nflationary  {R}elativity (\textsc{VIZIR}) which will also be made open-access in the future.

Typical operations of this package include: calculations of means, contrasts, and NR diagnostics but also power spectra, PDFs, and other statistical measures. In particular, one will find it useful to use BSSN to SVT formalism conversions such as in Appendix \ref{app:dicBSSN}, or vice-versa in Appendix \ref{app:extract}.

One of the key problems in physical extractions from numerical relativity is that of the impeded interpretation due to gauge ambiguities. If one is interested in a specific observer's point of view, it is not always guaranteed to be simulation-friendly given previous considerations on well-behaved numerical gauges. The usual gauges of NR are also rarely any close to observational ones. For these reasons, it is essential to extract gauge-invariant quantities such as ones built in perturbation theory (${\cal R}$, Bardeen quantities, etc ). However, that is only valid if the simulation stays perturbative and is not our purpose. In previous studies of non-perturbative cosmological frameworks, the key diagnostic extraction was that of the gauge-invariant matter energy density $\rho$ or more recently the Weyl curvature \cite{Ijjas2023,Elley2024} to include geometric contributions. We want to emphasize in our work the existence of a generalisation for the usual linear gauge-invariants such as ${\cal R}$.

The full GR comoving curvature on comoving hypersurfaces is defined as  \cite{rigopoulos_non-linear_2005}
\begin{equation}
    {\cal R}_i = -\frac{1}{6}\partial_i\ln |\gamma|+\frac{1}{6}\frac{\partial_0\ln |\gamma|}{\partial_0 \phi} \partial_i \phi,
\end{equation}
and constitutes a gauge-invariant quantity, even beyond the long-wavelength approximation \cite{langlois_evolution_2005}.\footnote{This was proved in \cite{langlois_evolution_2005} using the fluid 4-velocity as time direction instead of the normal observer's.} Its link with ${\cal R}$ becomes apparent when linearised
\begin{equation}
     {\cal R}_i^{(1)}  = {\cal R}_{,i}-\Phi_{B}(t_0)_{,i}-\frac{1}{3}\nabla^2E_{,i},
   \label{eq:RNLlin}
\end{equation}
which is the case in the geodesic gauge we work in (see Appendix \ref{app:RNL} for derivations). When running simulations so that all modes cross the horizon, terms other than ${\cal R}_{,i}$  are gradient-suppressed. This means that we now have a good probe for non-perturbativity and non-linearity by comparing ${\cal R}^{NL}_k = i\sum_a \frac{k_a}{k^2}({\cal R}_a+\Phi_B(t_0)-\frac{1}{3}\nabla^2E)$ and ${\cal R}$.

\begin{figure}
    \centering
\includegraphics[width=\linewidth]{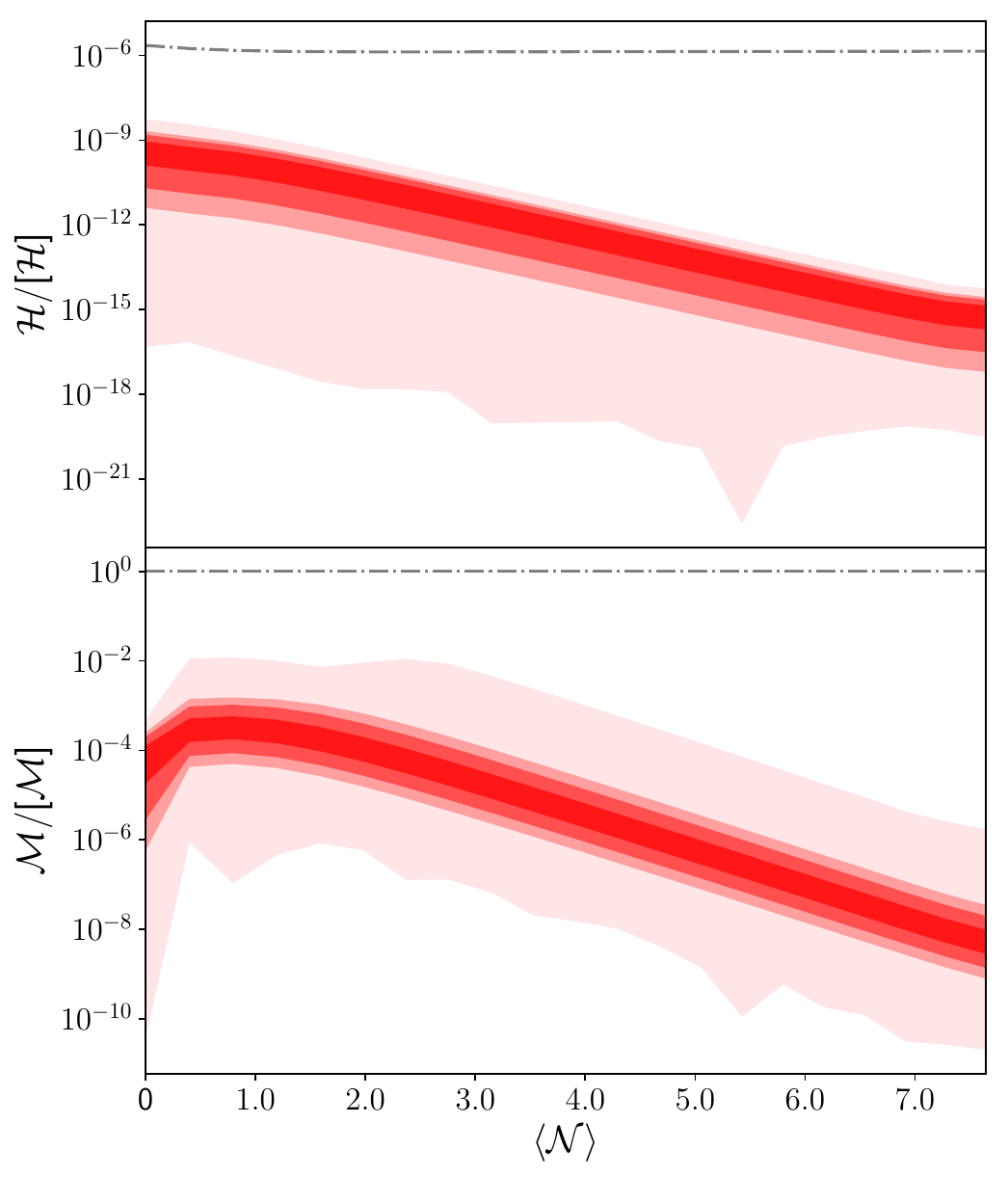}
    \caption{\justifying Distribution of the relative constraint violations over the box at each time step for quadratic inflation, using $68$, $95$, $99$ and $100$ percentile contours (dense red to transparent red). $[{\cal H}]$ and $[{\cal M}]$ are the usual sums of absolute respective terms in the constraints, see  \eqref{eq:abscons}. The dash-dotted line displays the average first-order hamiltonian and momentum constraint magnitudes $\langle[{\cal H}^{(1)}]\rangle/\langle[{\cal H}]\rangle$ and $\langle[{\cal M}^{(1)}]\rangle/\langle[{\cal M}]\rangle = 1$.}
\label{fig:quadviolations}
\end{figure}

\begin{figure}
\centering
\includegraphics[width=\linewidth]{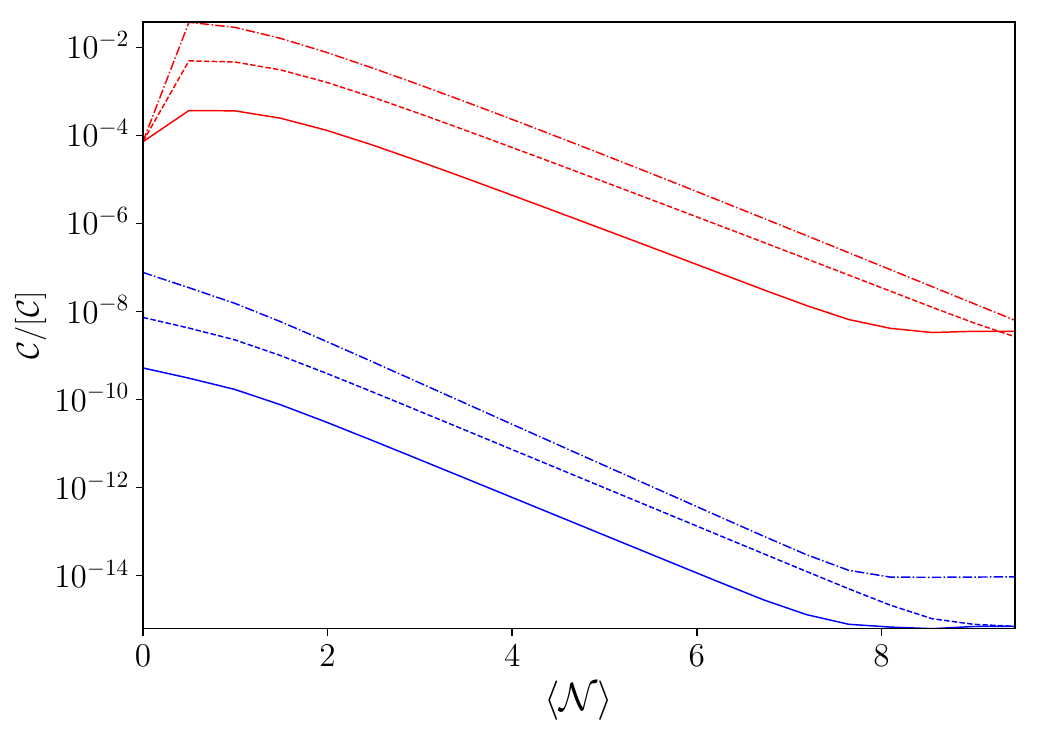}
\caption{ \justifying Convergence tests for the mean hamiltonian (blue) and momentum (red) relative constraints ($\langle|{\cal C}|/[{\cal C}]\rangle$, ${\cal C }= {\cal H}, {\cal M}$) across multiple runs with $N=64$ (dash-dotted), $N=128$ (dashed), and $N = 256$ (solid). Note that the stochastic initial conditions are the same ($\sigma = 3$).}
\label{fig:ConvTests}
\end{figure}

\subsection{Perturbative correctness and convergence \label{sec:correct}}

To evaluate the correctness of our simulations, a reference case or regime is required. For comparison, theoretical results from QFTCS are mostly available in slow-rolling perturbative scenarios of inflation.

In that respect, the background solution to the Friedmann equations and the linear prediction for the power spectrum from eq.~\eqref{eq:MS} will constitute a good diagnostic of correctness, especially because they can be solved using \textsc{STOIIC}, for the entire duration of the GR simulation.

Our vanilla case of inflation is the fully perturbative quadratic potential one $V(\phi) = \frac{1}{2} m^2\phi^2$ with $\phi_0$ and $m$ set using the \textit{potential-position fixing} method described by eq. \eqref{eq:potentialposition} and by imposing a slow-roll $\varepsilon_1 = -0.5\varepsilon_2 = 0.01$ and an initial Hubble constant of $H_0 = 10^{-5}M_{Pl}$. These values correspond to the BD vacuum fixing time and then are evolved for $\Delta {\cal N} = 5$ efolds until the start of the GR simulation.

Figure \ref{fig:meanquad} shows the background fields of interest from both the Friedmann equations and the GR simulation. The energy density is rendered at different times in Figures \ref{fig:3Da}, \ref{fig:3Db}, and \ref{fig:3Dc}. Figure \ref{fig:specquad} shows the initial and final dimensionless power spectra. The plots are cut at the value of $k_{max}^{\cal R}$ which is also where the initial spectrum is windowed with a Heaviside function. For this simulation, $\sigma = 3$ and $L = 64/H^{\circledast}$, hence a box satisfying our requirements of eq.~\eqref{eq:designbox}: slightly sub-Hubble and super-Hubble scales dynamics. 

The fit for the background quantities is perfect and suggests that inflation is proceeding correctly. On the inhomogeneous side, the superHubble freeze is visible for the density from Figure \ref{fig:3Db} to \ref{fig:3Dc}, and for the contrast of ${\cal R}$ in Figure \ref{fig:quad_rlin}. The extracted power spectra provide a very good fit to the linear prediction and affirm that ${\cal R}^{NL}$ is very close to ${\cal R}$ in a very perturbative case such as this one. The spectral turnover is also visible by eye from Figure \ref{fig:3Da} to \ref{fig:3Db}, where the clustering switches to smaller scales. These observations are of course all up to both cosmic and the estimator's variance, but also to higher order interactions, which, in this vanilla and barely sub-Hubble case, are very much negligible. 

From a full GR point of view, and as section \ref{sec:startingspacetime} emphasised, the constraints also need to be perfectly satisfied  as a key achievement in the construction of realistic initial conditions for NR. We define  the \textit{absolute constraint magnitudes} $[{\cal H}]$ and $[{\cal M}]$, calculated by summing the absolute value of the terms in the BSSN version of the constraints in eq. \eqref{eq:cons_eq}
\begin{equation}
\left\{
    \begin{aligned}
[{\cal H}] & \equiv  \big|{ }^{3}R\big|+\big|\frac{2}{3} K^2\big|+\big|\tilde{A}_{i j} {A}^{i j}\big|+\big|2 M_{Pl}^{-2} \rho\big|, \\
[{\cal M}_i] & \equiv \big|\tilde{\gamma}^{kl}  \partial_k\tilde{A}_{il} + 2\tilde{\gamma}^{kl}\tilde{\Gamma}^m_{l(i}\tilde{A}_{k)m}\big| \\
& +\big| 3\tilde{\gamma}^{kl}\tilde{A}_{ik}\frac{\partial_l\chi}{2\chi}\big| +\big|\frac{2}{3} K_{\mid i}\big|+\big|M_{Pl}^{-2} {\cal { J}}_i\big|,\\
[{\cal M}] & \equiv \sqrt{\sum_i [{\cal M}_i]^2}.     \end{aligned}\right .
    \label{eq:abscons}
\end{equation}
{These will typically show the magnitude of the different individual terms before cancellation and  we require our simulations to have $|{\cal H}|/[{\cal H}]\ll 1$ and $|{\cal H}|/[{\cal H}]\ll 1$ as a measure of how well the constraints are satisfied.} Note that these magnitudes are different in the case of inhomogeneous inflation: $[{\cal H}]$ is usually of order $H^2$ coming from the background, while $[{\cal M}]$ has a leading term at linear order of the perturbations since the Friedmann homogeneous universe trivially satisfies the momentum constraint. This implies that if our constraints are satisfied at linear order, $|{\cal H}|/[{\cal H}]$ should show the ratio of two orders of perturbation theory while only one for $|{\cal M}|/[{\cal M}]$. Note that it is still possible to look at the gap between our hamiltonian violation and the first order magnitude by computing $[{\cal H}^{(1)}]/[{\cal H}]$.

Figure \ref{fig:quadviolations} shows the box distribution of these two ratios at each time step. We observe a satisfaction much below the linear magnitude, hence a violation at second order in perturbation theory initially. However, a natural damping of this violation occurs, mostly likely thanks to the expansion scaling. This means that the higher-order interactions within our box, here at second order, can be safely investigated. We will see that these conclusions can be different when reaching non-perturbative regimes.

Finally, to check that the error built during the evolution is numerical and that it reduces appropriately when increasing resolution, one performs convergence tests. In NR, the constraints are once again useful here because of their direct link to the physicality of the solution. In this case, the scaling one should observe for the constraint ${\cal C}(N)$ from a 3D discrete simulation with $N\gg 1$ is \cite{radia_lessons_2022}
\begin{equation}
    \frac{{\cal C}(N_1)}{{\cal C}(N_2)} = \left ( \frac{N_2}{N_1}\right )^n,
\end{equation}
where $n$ is the spatial order of convergence of the stencils used, $n=4$ in our \textit{GRChombo} example.

Figure \ref{fig:ConvTests} gives the right scaling ($\simeq (\frac{128}{256})^4\simeq 0.06$) in the evolution for the $N = 128$ to $N=256$ transition. Initial constraints can be overlooked as they are built from our linear solutions and so will have higher order violations in perturbation theory, mixed with some of the negligible numerical errors we are studying the scaling of. Final scalings can become uninformative too as the violation reaches the bottom of numerical accuracy and stagnates.

\section{Non-perturbative examples \label{sec:Applications}}
In the following, we study models known for their non-perturbative properties using full GR, providing tests of our pipeline against { more extreme scenarios compared to the slow-roll case studied above}.

\begin{figure}
        \centering
\includegraphics[width=0.95\linewidth]{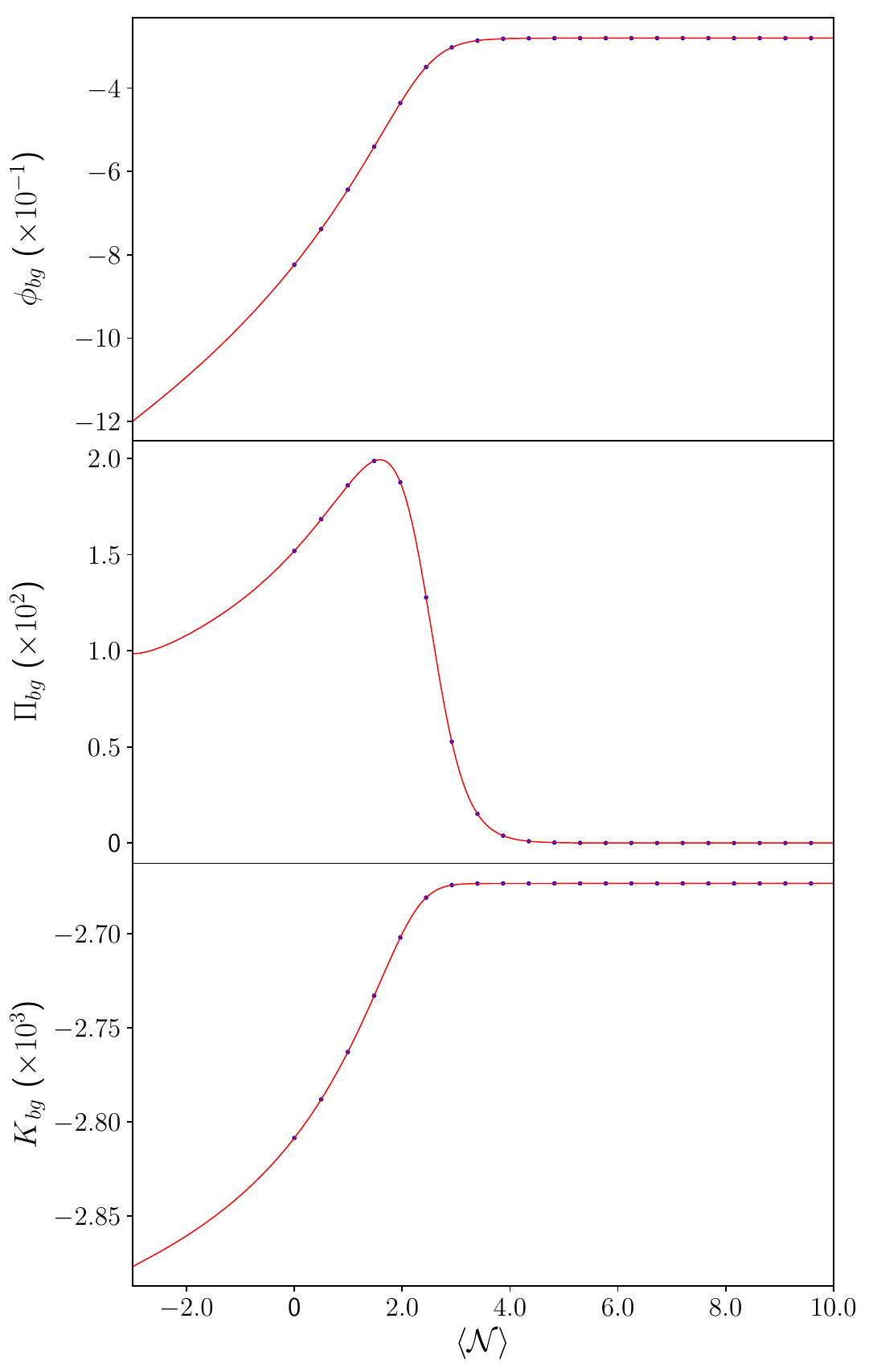}
\caption{\justifying Comparison of the Friedmann solution (solid red) and the mean inflaton  (top), conjugate momentum (middle), and extrinsic curvature (bottom) fields extracted from the GR simulation (blue dots) in inflection inflation. $M_{Pl}=100$ units.}
\label{fig:meanUSR}
\end{figure}

\subsection{Inflection point inflation \label{subsec:inflection}}
We study a single field rolling on the following potential
\begin{equation}
V(\phi) = \Lambda v^4 \frac{\phi^2}{3} \cdot \frac{3\phi^2 + 2\sqrt{2}\phi v + 6v^2}{(3\phi^2 + 2v^2)^2},
\label{eq:potProkopec}
\end{equation}
inspired by Higgs inflation \cite{Hamada14,Bezrukov14} and intended to be tuned for the production of primordial black holes (PBHs) in \cite{GarciaBellido17,germani17}. It produces three phases around the inflection point at $\phi_{inf}\simeq 0.28M_{Pl}$: ultra slow-roll (USR) sandwiched by slow-roll phases. The USR phase is interesting because it provokes a resonant enhancement of the adiabatic power spectrum. In \cite{germani17}, parameters are tuned to have that peak on sub-CMB scales but we slightly change $v$ to see interesting dynamics within $10$ efolds. 

\begin{figure}[h]
    \centering
\includegraphics[width=0.95\linewidth]{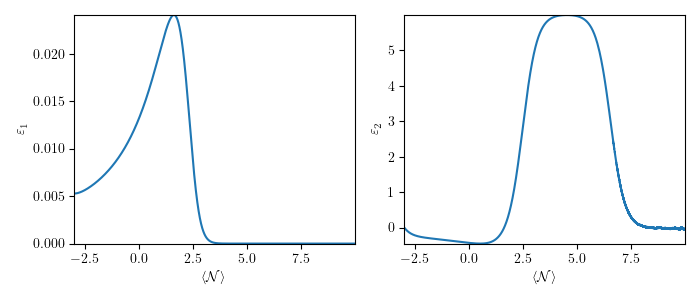}
\caption{\justifying First (left) and second (right) slow-roll parameters for the inflection inflation example.}
\label{fig:usrrolls}
\end{figure}

Our parameters are $\Lambda = 1.86\times 10^{-6}$, $v = 0.198M_{Pl}$ and we start slow-rolling before the feature with $\phi_0 = -1.2M_{Pl}$ using the \textit{$H_0$-fixing} method described in section \ref{sec:startingspacetime}. These parameters give a decent duration of USR ($\varepsilon_1\simeq 0$, $\varepsilon_2\simeq 6$, yielding a field that almost stops) as Figures \ref{fig:meanUSR} and  \ref{fig:usrrolls} showcase. 

The box is designed as follows: modes up to $\sigma = 4$ are drawn and accounted for in a patch of size $L=32/H^{\circledast}$. The resulting background of the simulation in Figure \ref{fig:meanUSR} matches the Friedmann prediction perfectly.

\begin{figure}[h]
    \centering
\includegraphics[width=0.9\linewidth]{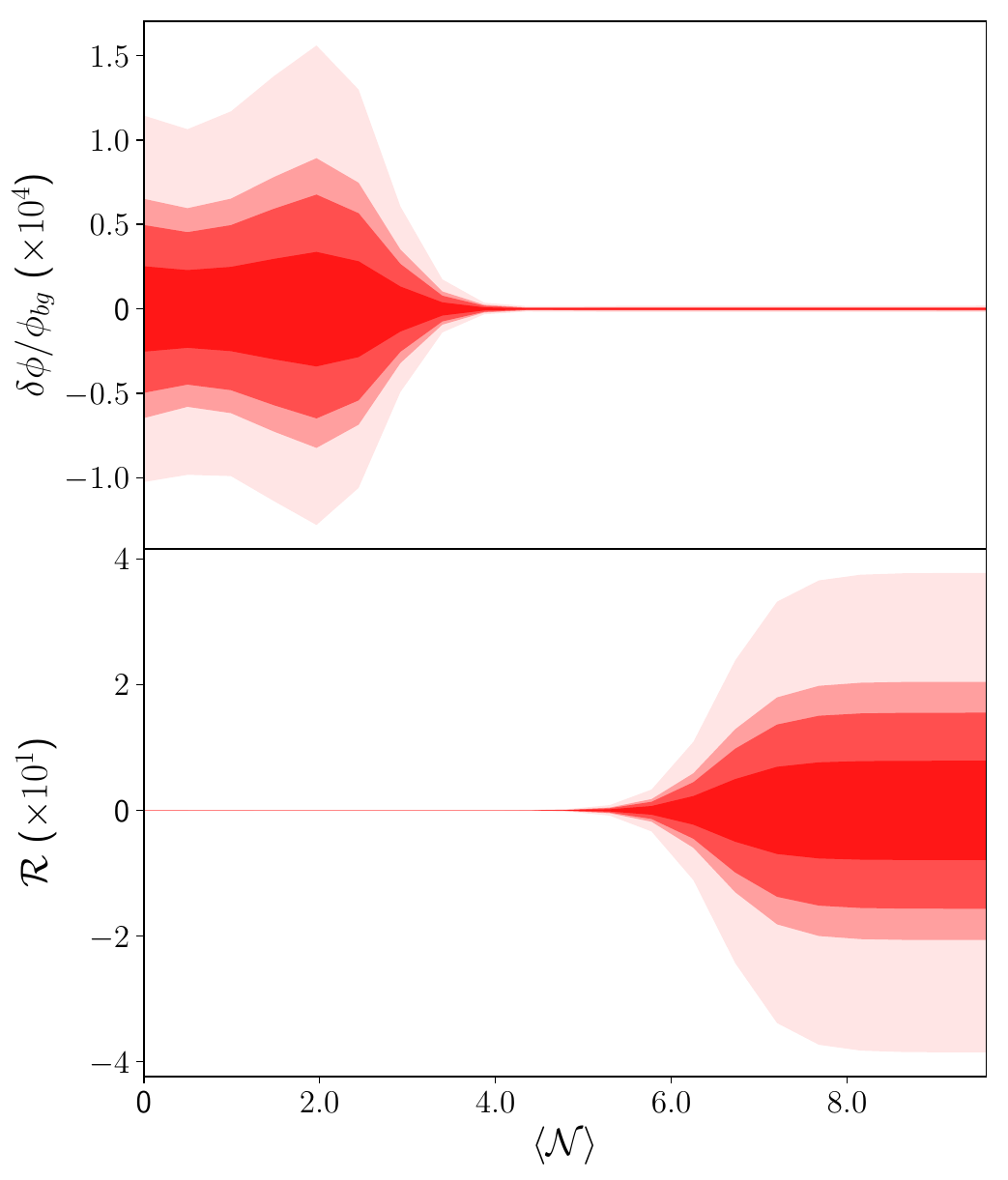}
    \caption{\justifying Contrast of $\phi$ and ${\cal R}$ during inflection inflation using $68$, $95$, $99$ and $100$ percentile contours (dense red to transparent red). $M_{Pl}=100$ units.}
    \label{fig:usr_rlin}
\end{figure}

\begin{figure*}
    \centering
\includegraphics[width=0.7\linewidth]{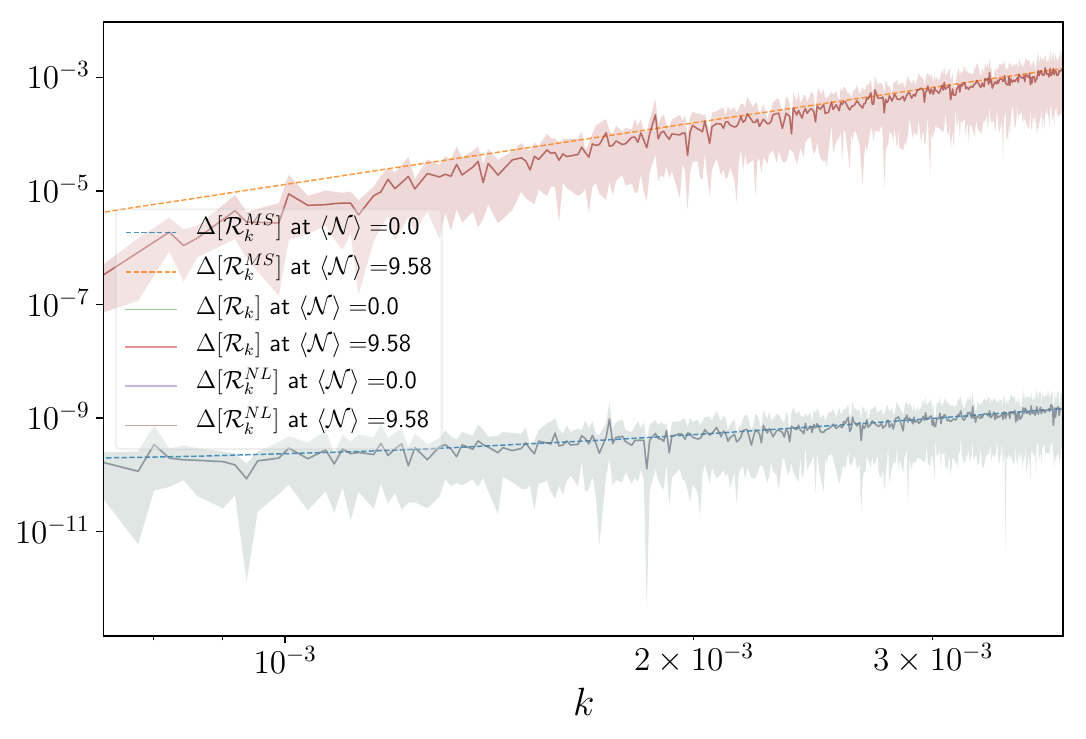}
\caption{\justifying Linear (dashed, ${\cal R}^{MS}$) and extracted (solid, ${\cal R}$) binned dimensionless spectra ($\Delta[{\cal R}]= \frac{k^3}{2\pi^2} |{\cal R}_k|^2$) compared to extracted ${\cal R}^{NL}$, scalar built from the fully non-linear generalisation ${\cal R}_i$ in the inflection inflation case{: perfect superimposition at initial (lower half) and final (upper half) times ({\textcolor[HTML]{95d095}{\rule{1em}{0.5em}}} $+$ {\textcolor[HTML]{cab3de}{\rule{1em}{0.5em}}} $=$ \textcolor[HTML]{608475}{\rule{1em}{0.5em}}, {\textcolor[HTML]{ea9393}{\rule{1em}{0.5em}}} $+$ {\textcolor[HTML]{c6aba5}{\rule{1em}{0.5em}}} $=$ \textcolor[HTML]{b13e3a}{\rule{1em}{0.5em}})}. Contours account for the $3\sigma$-range of points in each bin. Displayed Fourier bins are those of the simulation mesh which have more than 10 realisations and which are smaller than the spectral cutoff of the input ($k_{max}^{\cal R} = 4\times 10^{-3}$). $M_{Pl}=100$ units.}
\label{fig:specUSR}
\end{figure*}

This example is the first to showcase a non-perturbative solution. The contrasts can be summarised in that of ${\cal R}$, showing a departure from perturbativity in Figure \ref{fig:usr_rlin}: up to a $30\%$ difference from the background. One would thus expect a consequent departure from linear theory and perturbation theory in general.
The spectrum is plotted in Figure \ref{fig:specUSR}. The simulation shows indeed  {a small deviation from} linear theory compared to the quadratic case {at low $k$}. However, it shows no visible difference between ${\cal R}$ and ${\cal R}^{NL}$, despite the complicated and less perturbative interactions occurring at the inflection's phase transition.

\begin{figure*}
    \centering
         \begin{subfigure}{0.32\textwidth}
        \centering
\includegraphics[width=\linewidth]{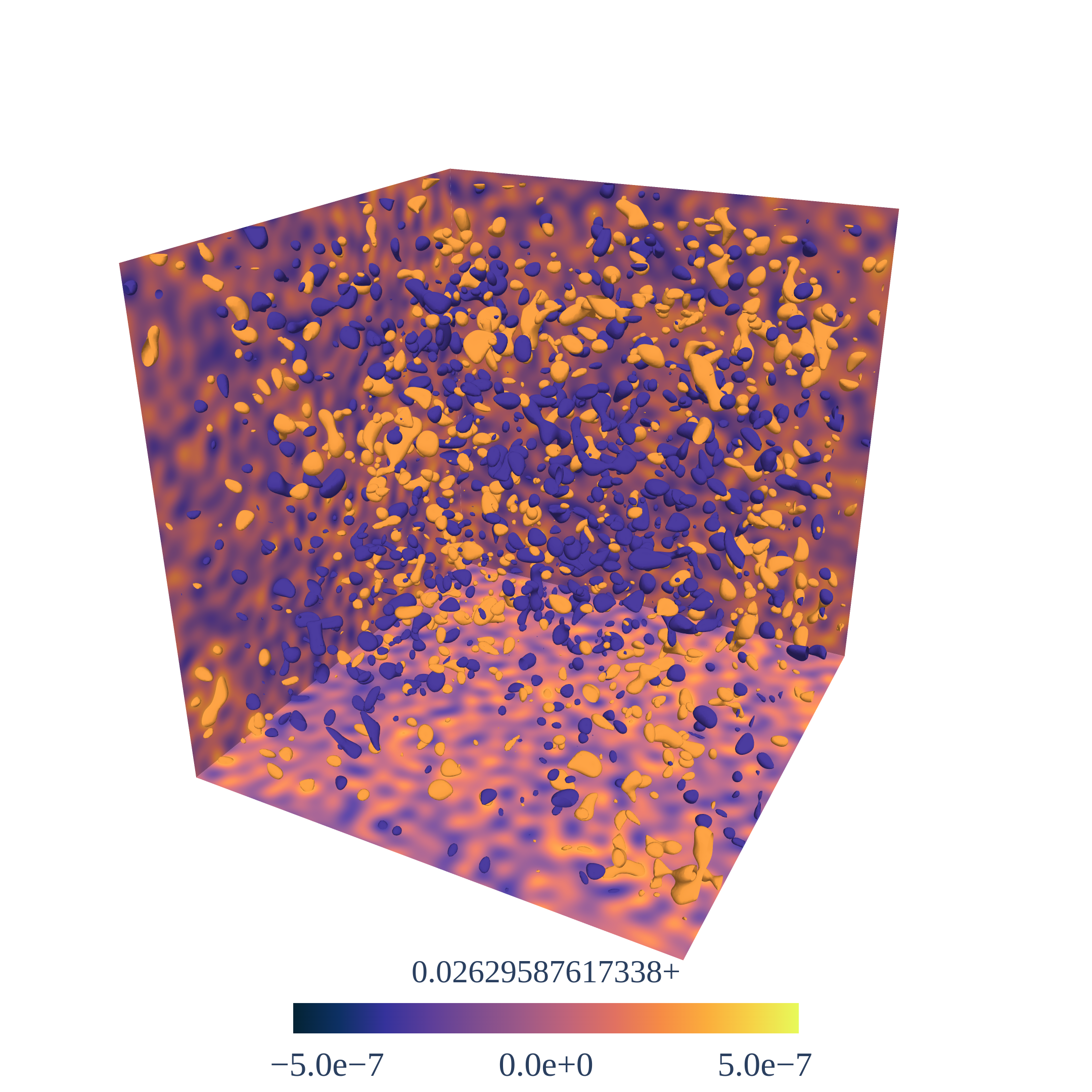}
    \caption{Inflection at $\langle {\cal N}\rangle=0$}
    \label{fig:3Dd}
    \end{subfigure}
    \hfill
\begin{subfigure}{0.32\textwidth}
        \centering        \includegraphics[width=\linewidth]{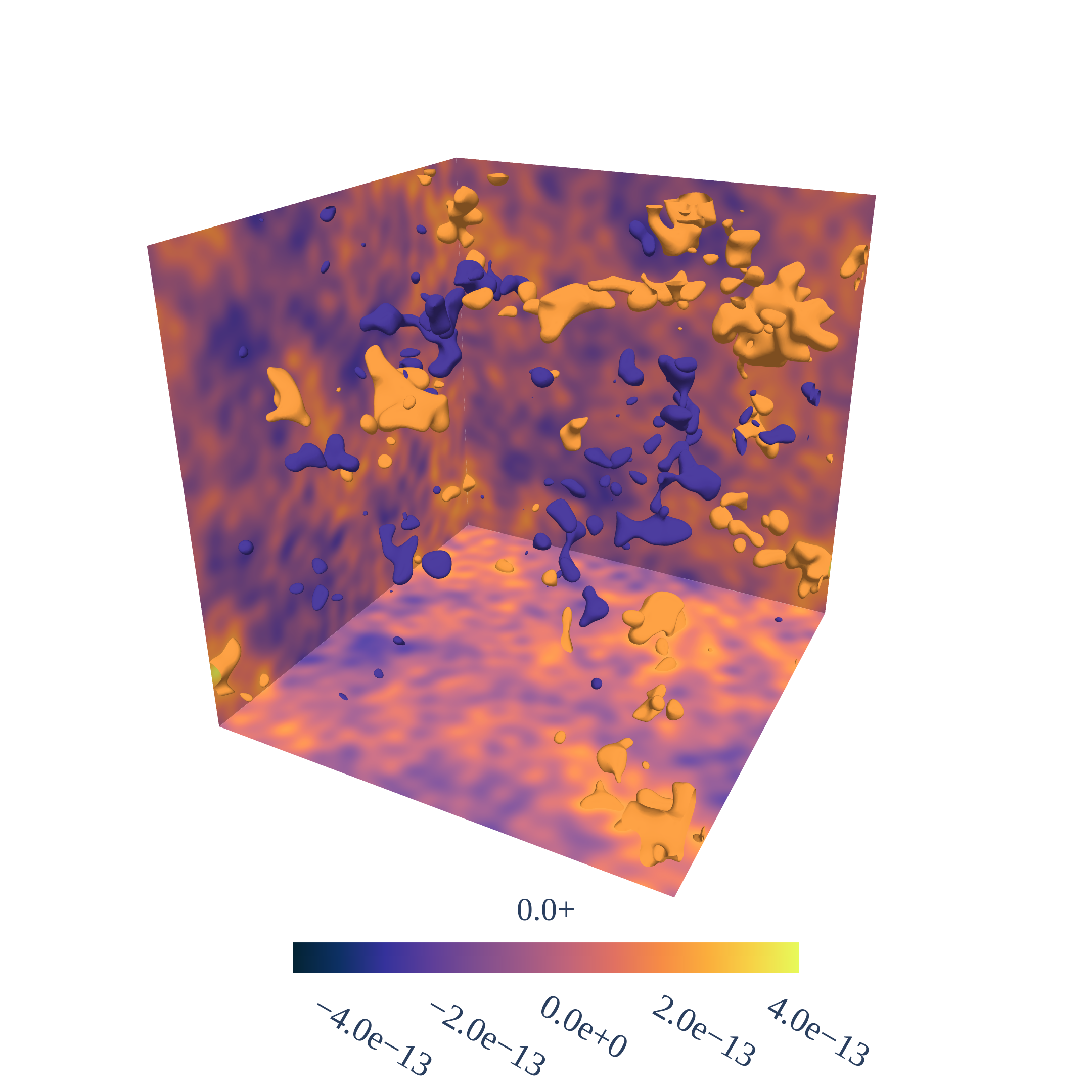}
        \caption{Inflection at $\langle {\cal N}\rangle=4.82$}
        \label{fig:3De}
    \end{subfigure}
\hfill
\begin{subfigure}{0.32\textwidth}
        \centering
        \includegraphics[width=\linewidth]{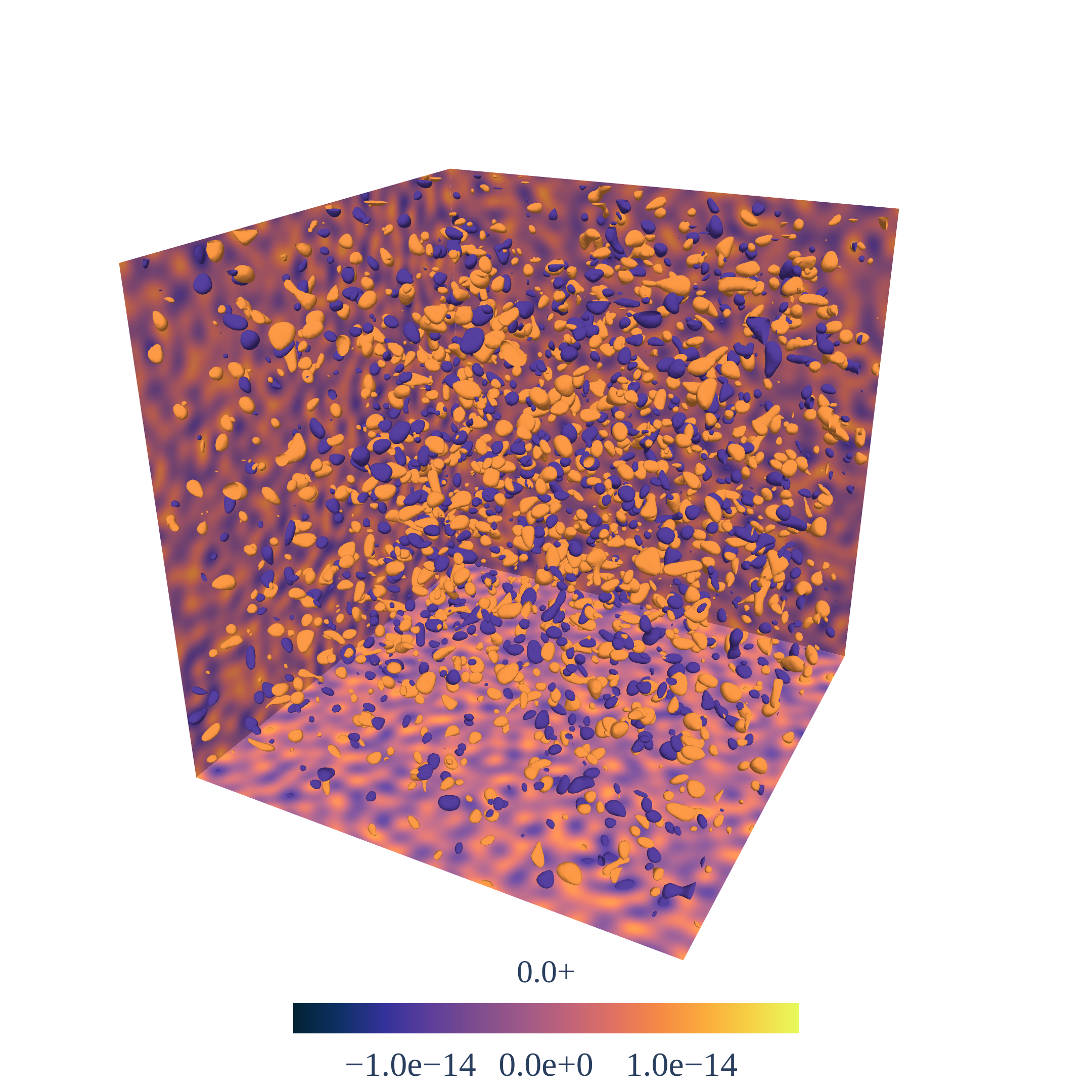}
        \caption{Inflection at $\langle {\cal N}\rangle=9.58$}
        \label{fig:3Df}
    \end{subfigure}

    \centering
        \begin{subfigure}{0.32\textwidth}
        \centering
\includegraphics[width=\linewidth]{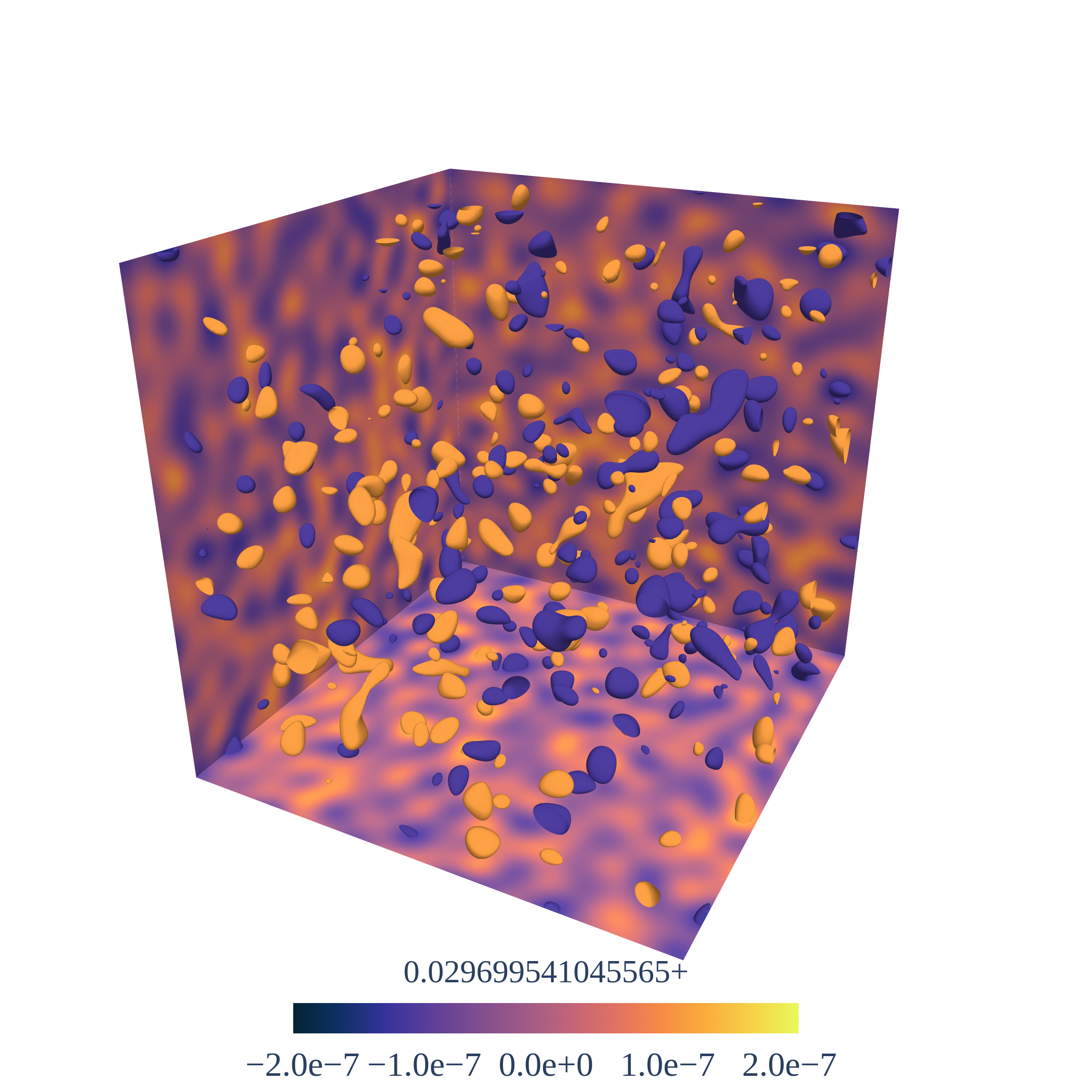}
    \caption{Strong resonance at $\langle {\cal N}\rangle=0$}
    \label{fig:3Dg}
    \end{subfigure}
    \hfill
\begin{subfigure}{0.32\textwidth}
        \centering        \includegraphics[width=\linewidth]{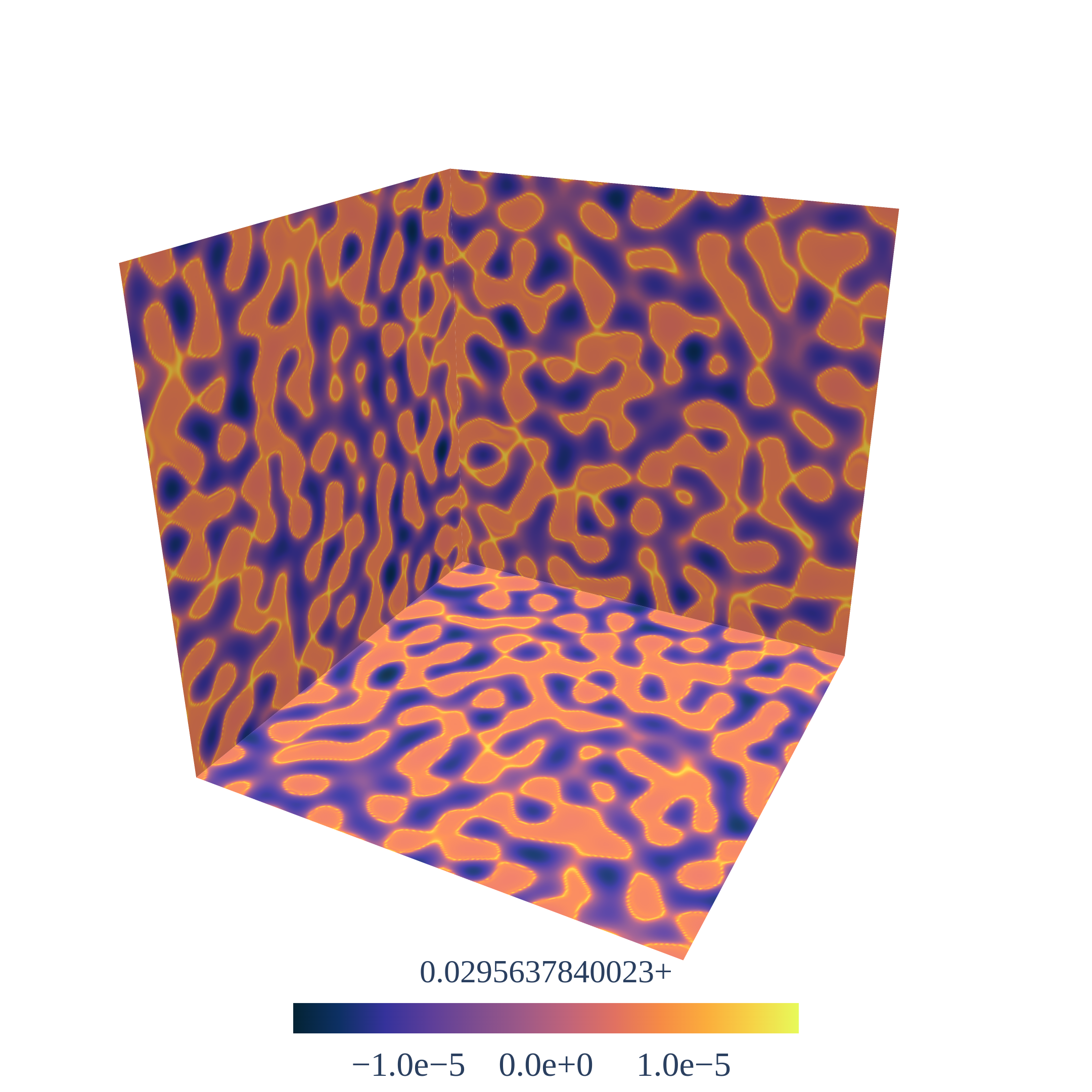}
        \caption{Strong resonance at $\langle {\cal N}\rangle=3.83$}
        \label{fig:3Dh}
    \end{subfigure}
\hfill
\begin{subfigure}{0.32\textwidth}
        \centering
        \includegraphics[width=\linewidth]{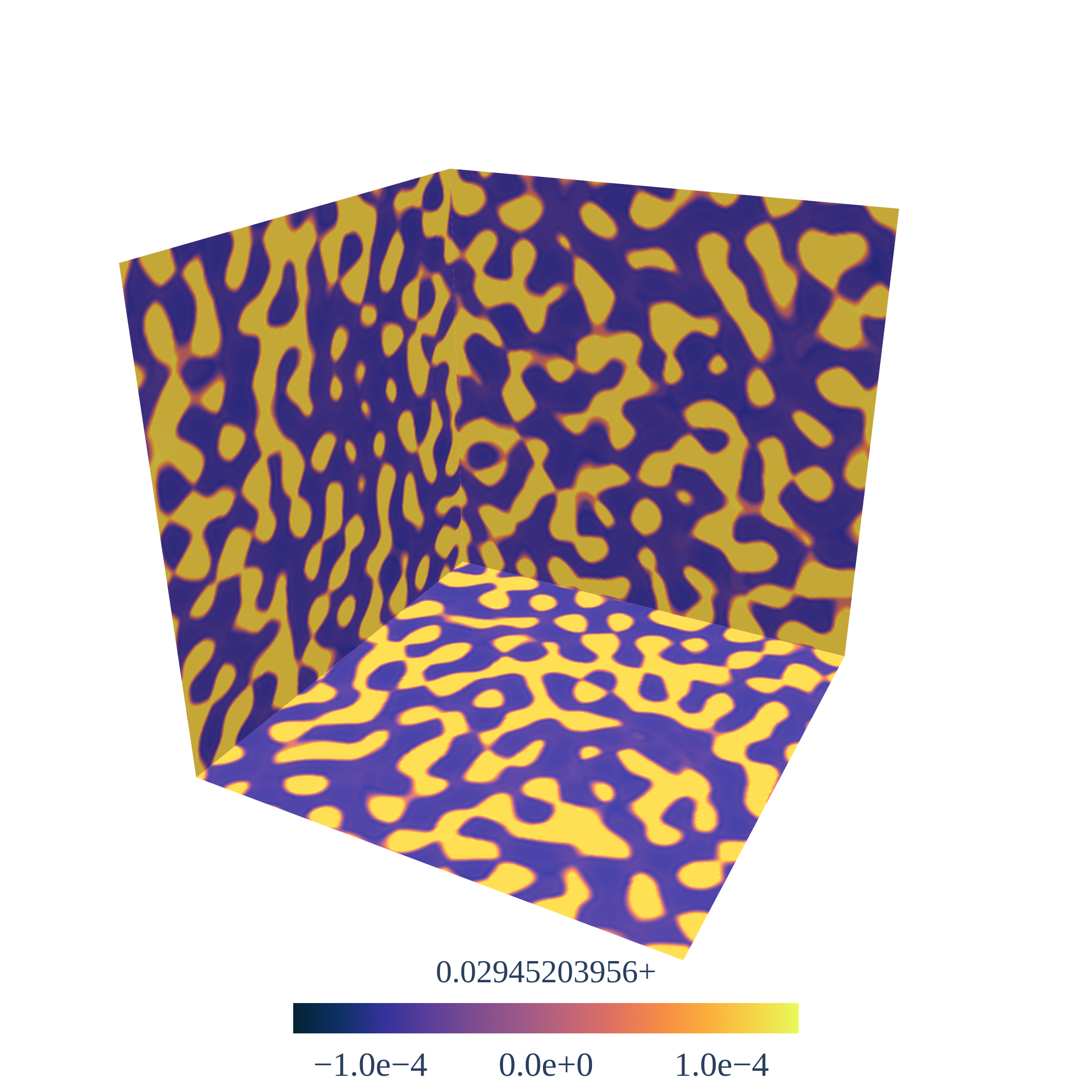}
        \caption{Strong resonance at $\langle {\cal N}\rangle=6.98$}
        \label{fig:3Di}
    \end{subfigure}
\caption{\justifying Energy densities evolution (left to right) of the simulated patches for the two non-perturbative studied models (top to bottom). 2D boundary heat maps and volumic isosurfaces share the same color bar, but not across plots. Isosurfaces are at $\pm 2.5$std for the inflection example. Note that the resonance is not shown with any isosurface for better visibility. $M_{Pl}=100$ units. }
\label{fig:3Dplots}
\end{figure*}

Interestingly, the {field} contrasts shrink greatly during the USR phase but benefit that of ${\cal R}$ through the geometry, see Figure \ref{fig:usr_rlin}. The drop in $\delta\phi$ appears analogous to what the long wavelength and separate universe Hamilton-Jacobi theory predicts for motion on a plateau \cite{prokopec_ensuremathdeltan_2021}, showing consistency with that idealized scenario, as well as the similar decay shown in \cite{Prokopec:2025uvz}.
The evolution of the energy density is also illustrated in Figures \ref{fig:3De}, \ref{fig:3Df}, and \ref{fig:3Dg} and shows that the clustering has been altered during the phase transition as opposed to the quadratic case.

\begin{figure}[h]
    \centering
\includegraphics[width=0.9\linewidth]{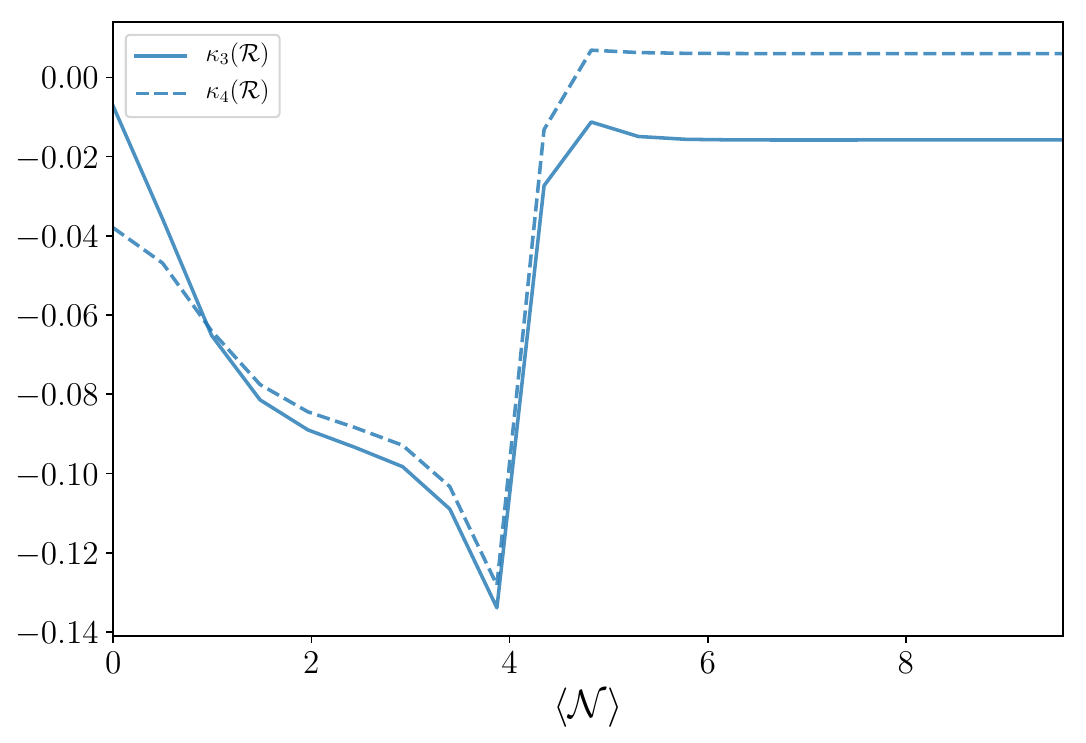}
    \caption{\justifying Cumulants of ${\cal R}$ from the GR simulation of the inflection inflation: sample skewness ($\kappa_3$, solid blue) and kurtosis ($3+\kappa_4$, dashed blue).}
    \label{fig:cumulUSR}
\end{figure}

Most importantly, such scenarios are of interest to those studying primordial non-Gaussianity. In Figure \ref{fig:cumulUSR} we report a consequent influence of the transition dynamics on the skewness and kurtosis. Leaving a more thorough study for future work, we refer to \cite{Florio2024} for another example in the tensorial sector.

\subsection{Strong resonance
\label{subsec:strongRes}}
\begin{figure}
    \centering
\includegraphics[width=0.9\linewidth]{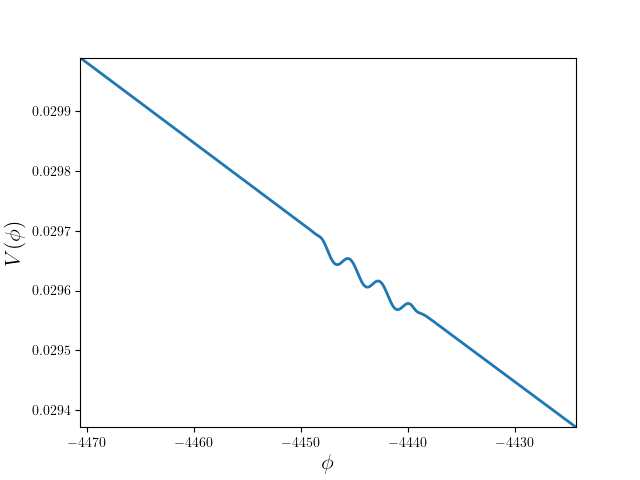}
\caption{\justifying Strong resonance potential: monodromic feature on the quadratic potential. The range in $\phi$ is the one explored during the simulations. $M_{Pl}=100$ units.}
\label{fig:potmono}
\end{figure}

Resonance models have received significant attention as inflationary models with non-trivial features \cite{chen_generation_2008,Flauger10}.
They often consist of a usual vanilla potential, on top of which an oscillatory feature is added. Some models were tested against Planck \cite{PlanckInflation,martin_best_2014, Sohn24} without decisive conclusions. These were still perturbative as, once again, it is our only way to make predictions using QFTCS.

More recently, non-perturbative simulations of inflation gave their first results using lattice cosmology \cite{Caravano2024}, building up from reheating studies \cite{latticeeasy}. In this original study, the non-perturbative regime was reached using a strong resonance of the perturbations with the oscillatory feature, inspired by axion monodromy models \cite{Monodromy10}. Despite offering important insights, these simulations remain not fully relativistic and for instance, approximate backreaction and geometry, although some recent improvements clarified some regimes \cite{caravano2024axionbackreaction,caravano2024usr}. We believe that our framework can {illuminate} a complete relativistic picture of these processes.

We use a potential similar to that of \cite{Caravano2024}, which is a standard vanilla inflation supplemented by a monodromic feature of width $\Delta\phi$
\begin{equation}
\left\{
\begin{aligned}
   V(\phi)= &  \frac{1}{2}m^2\phi^2+\Lambda^4{\cal W}(\phi-\phi_c)\left(\cos\left(\frac{\phi-\phi_c}{f}\right)-1\right), \\
   {\cal W}(\phi) = & \frac{1}{4}\left(1+\tanh \left(\frac{\phi}{f}\right)\right)\left(1+\tanh \left(\frac{\Delta \phi-\phi}{f}\right)\right),
   \end{aligned}\right .
\end{equation}
the only difference being that our choice of slow-rolling background is taken to be the same as our vanilla example\footnote{We thank the authors for clarifying their initialisation methods. Here, we have chosen a different set of initial conditions instead, easier to implement within our framework.}. The oscillatory feature is located such that $\phi_c$ is the value of $\phi^{\circledast}$ in the quadratic example. Finally, we set our parameters to $\alpha = \dot{\phi}_c/fH_c = 10$, $b = 2\Lambda^4/\dot{\phi}_c^2 = 1.32$, and $\beta = \Delta\phi/\alpha f = 2$, respectively fixing frequency, amplitude, and duration of the oscillations. The resulting potential is shown in Figure \ref{fig:potmono}.

\begin{figure}
        \centering
\includegraphics[width=0.95\linewidth]{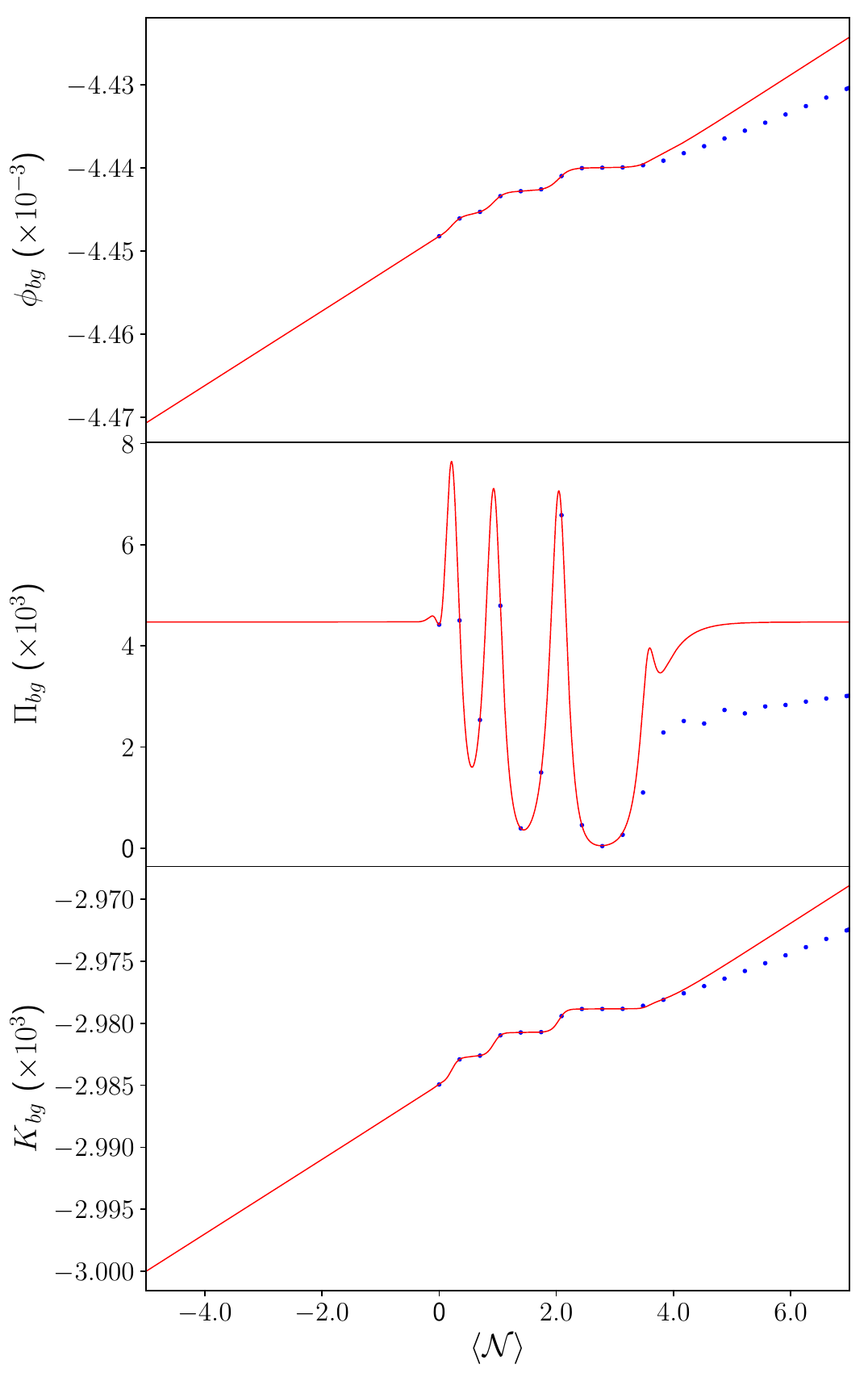}
\caption{\justifying Comparison of the Friedmann solution (solid red) and the mean inflaton (top), conjugate momentum (middle), and extrinsic curvature (bottom) fields extracted from the GR simulation (blue dots) in a strong resonance inflation. $M_{Pl}=100$ units. }
 \label{fig:meanmono}
\end{figure}

\begin{figure}
    \hspace*{-0.7cm}
\includegraphics[width=1\linewidth]{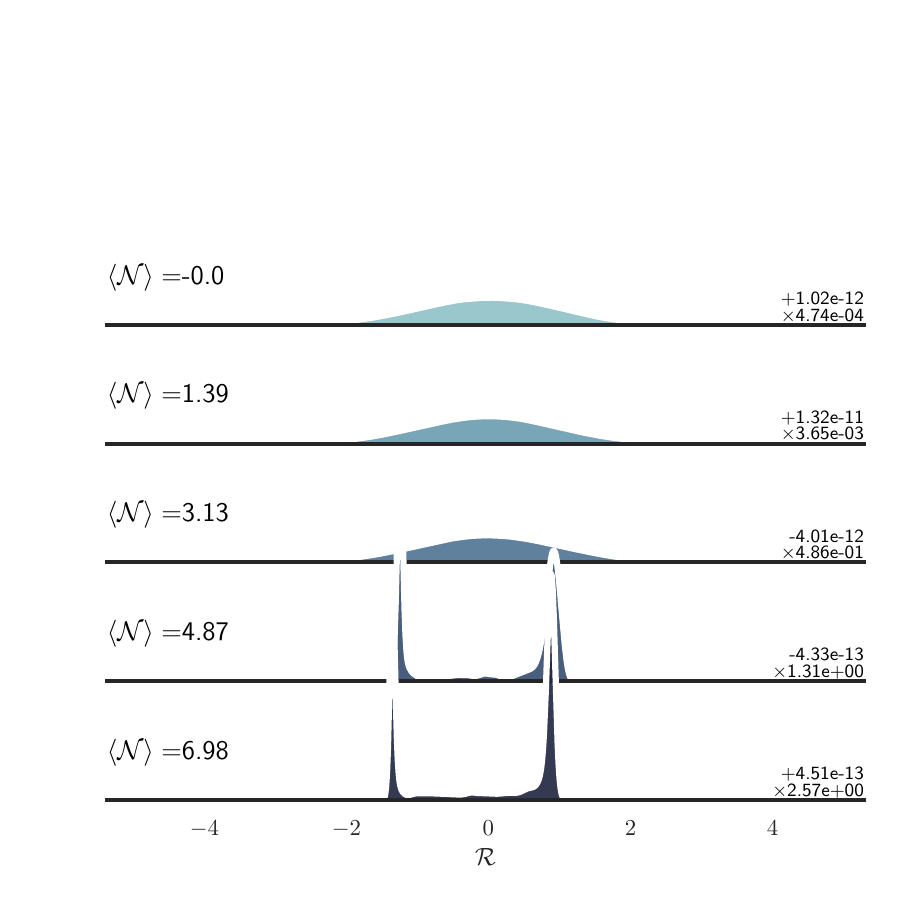}
    \caption{\justifying PDFs of ${\cal R}$ during the strong resonance evolution: from gaussianity to bimodal fluctuations.}
    \label{fig:monoPDFs}
\end{figure}

We start with a range of scales up to $\sigma = 4.5$ in a box of physical size $L = 16/H^{\circledast}$, necessary to probe slightly sub-Hubble scales to observe the resonance with this choice of parameters. In this highly non-perturbative scenario, the background loses its meaning as shown in Figure \ref{fig:meanmono}: compared to the two previous examples, measurable deviations from the Friedmann solution occur after a few efolds in the oscillating features, with those of $\Pi$ particularly pronounced. In particular, looking at the evolution of the energy density contrasts in Figure \ref{fig:3Dg}, \ref{fig:3Dh}, and \ref{fig:3Di} or at the PDFs in the Joy plot of Figure \ref{fig:monoPDFs} provides a very good grasp of the dynamics. It shows that the perturbations become bimodal and non-perturbative in the gap between the two groups. A more acute look at the values of the fields shows that some perturbations stay stuck in some of the bumps of the potential: these will undergo eternal inflation in a false vacuum while others make it through en route for successful inflation. These are the same qualitative conclusions as in \cite{Caravano2024}.

\begin{figure*}
    \centering
\includegraphics[width=0.7\linewidth]{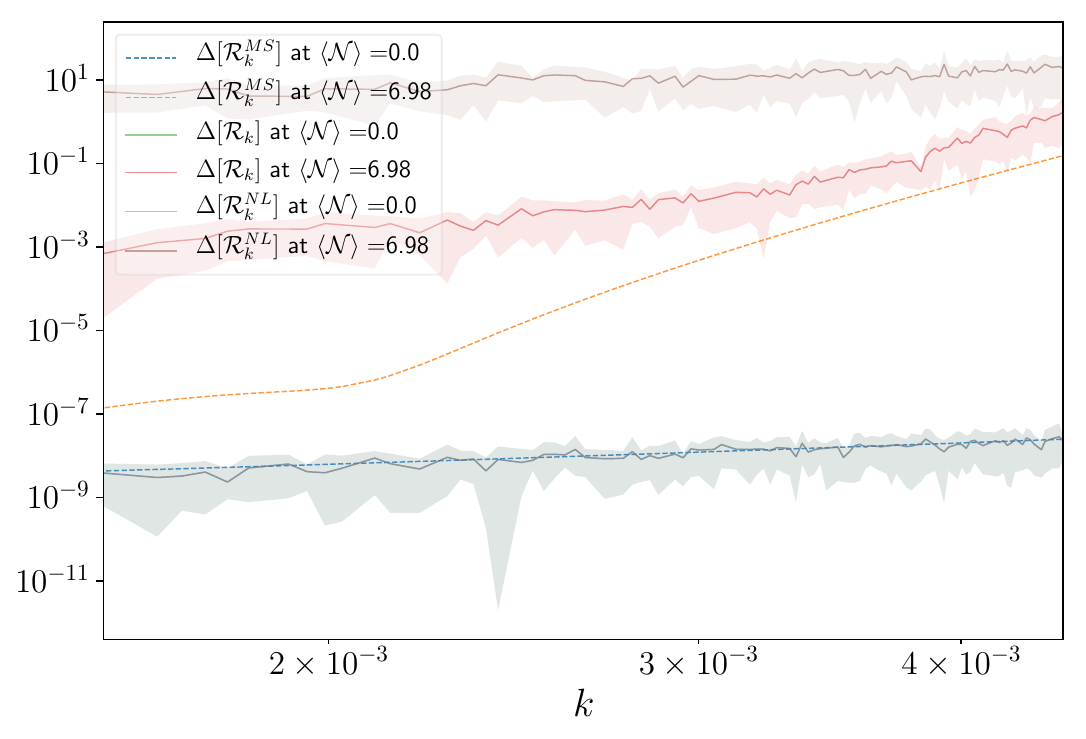}
\caption{\justifying  Linear (dashed, ${\cal R}^{MS}$) and extracted (solid, ${\cal R}$) binned dimensionless spectra ($\Delta[{\cal R}]= \frac{k^3}{2\pi^2} |{\cal R}_k|^2$) compared to extracted ${\cal R}^{NL}$, scalar built from the fully non-linear generalisation ${\cal R}_i$ in the strong resonance case{: perfect superimposition at initial (lower half) time only ({\textcolor[HTML]{95d095}{\rule{1em}{0.5em}}} $+$ {\textcolor[HTML]{cab3de}{\rule{1em}{0.5em}}} $=$ \textcolor[HTML]{608475}{\rule{1em}{0.5em}})}. Contours account for the $3\sigma$-range of points in each bin. Displayed Fourier bins are those of the simulation mesh which have more than 10 realisations and which are smaller than the spectral cutoff of the input ($k_{max}^{\cal R} = 4.5\times 10^{-3}$). $M_{Pl}=100$ units.}
\label{fig:specmono}
\end{figure*}

Moreover, we also found an encouraging resemblance between the cited study and our spectra in Figure \ref{fig:specmono}. The departure from linear theory is extreme and likewise, but mostly motivates one to stop looking at a perturbative quantity such as ${\cal R}$. ${\cal R}^{NL}$ shows indeed great independence from the former.

\subsection{Numerical discussion \label{subsec:comments}}

\begin{figure*}
    \centering
    \begin{subfigure}{0.49\textwidth}
\includegraphics[width=\linewidth]{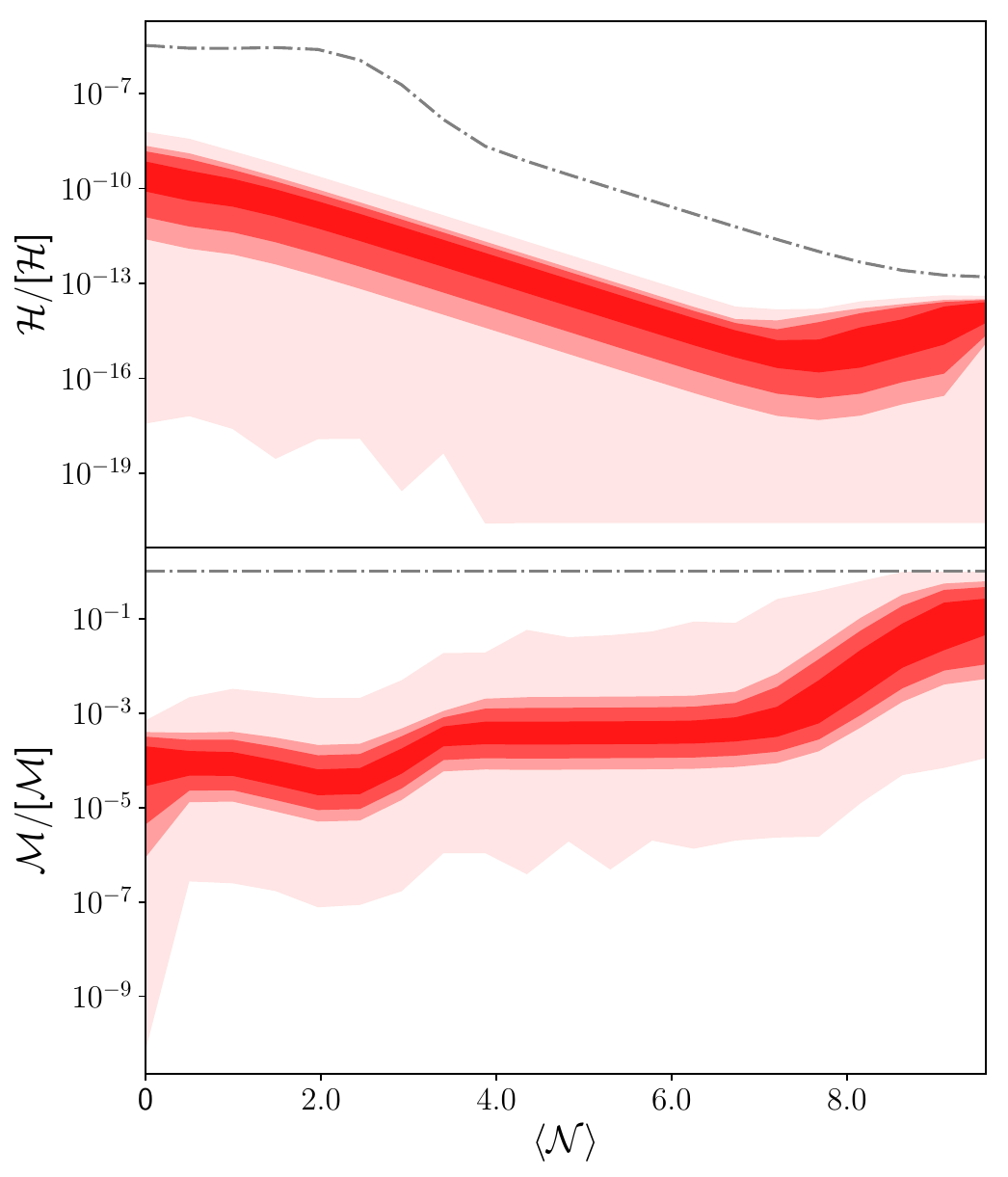}
\caption{Inflection inflation}
 \label{fig:NPconstraintsa}
    \end{subfigure}
    \begin{subfigure}{0.49\textwidth}
\includegraphics[width=\linewidth]{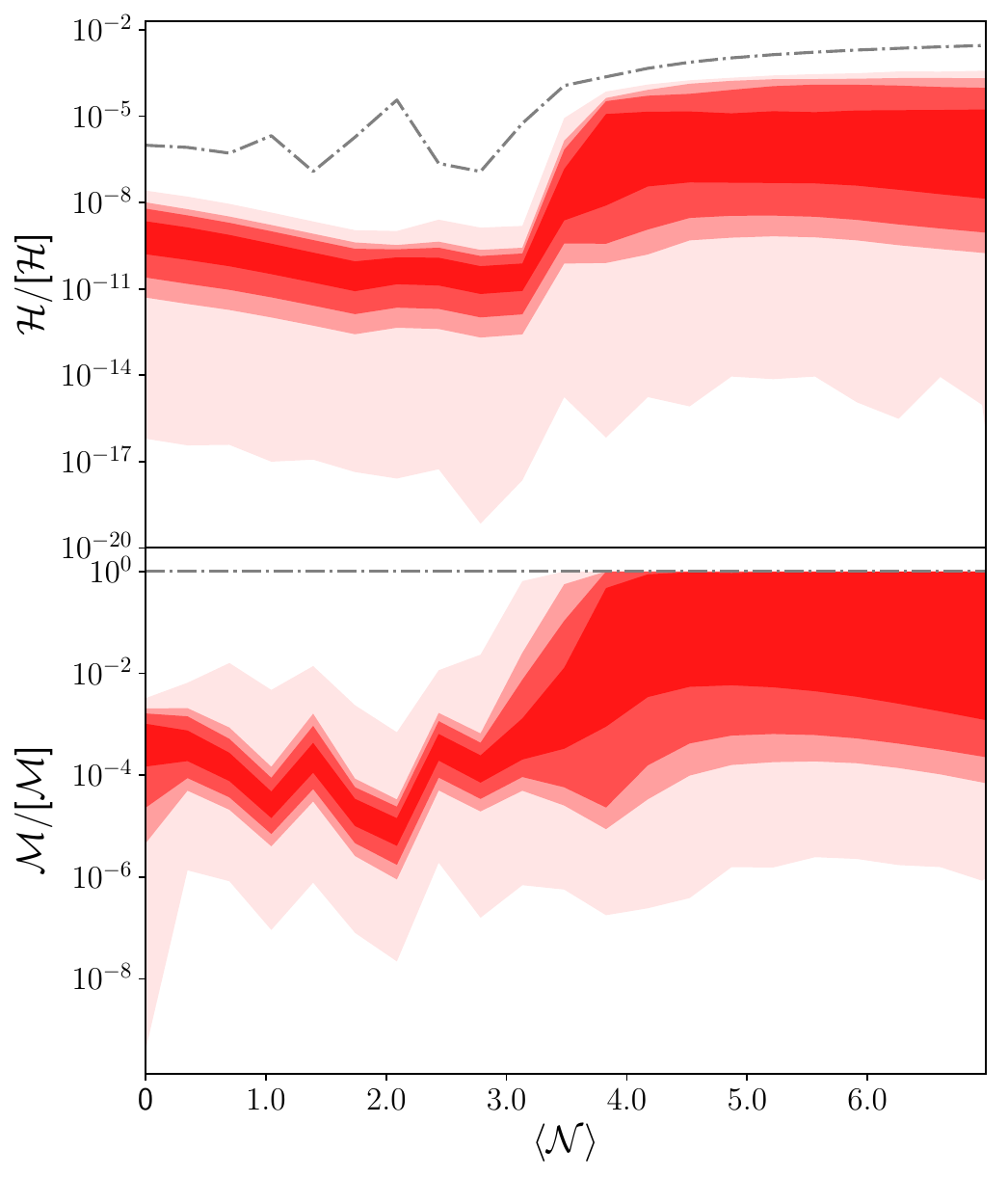}
\caption{Strong resonance }
 \label{fig:NPconstraintsb}
    \end{subfigure}
    \caption{\justifying Evolution of the rescaled constraint violation percentiles for our non-perturbative models. For each time step's box, $68$, $95$, $99$ and $100$ percentile contours are represented (dense red to transparent red). The dash-dotted line displays the average first-order hamiltonian and momentum constraint magnitudes $\langle[{\cal H}^{(1)}]\rangle/\langle[{\cal H}]\rangle$ and $\langle[{\cal M}^{(1)}]\rangle/\langle[{\cal M}]\rangle = 1$.}
    \label{fig:NPconstraints}
\end{figure*}

\begin{figure}[h]
    \hspace*{-0.7cm}
\includegraphics[width=\linewidth]{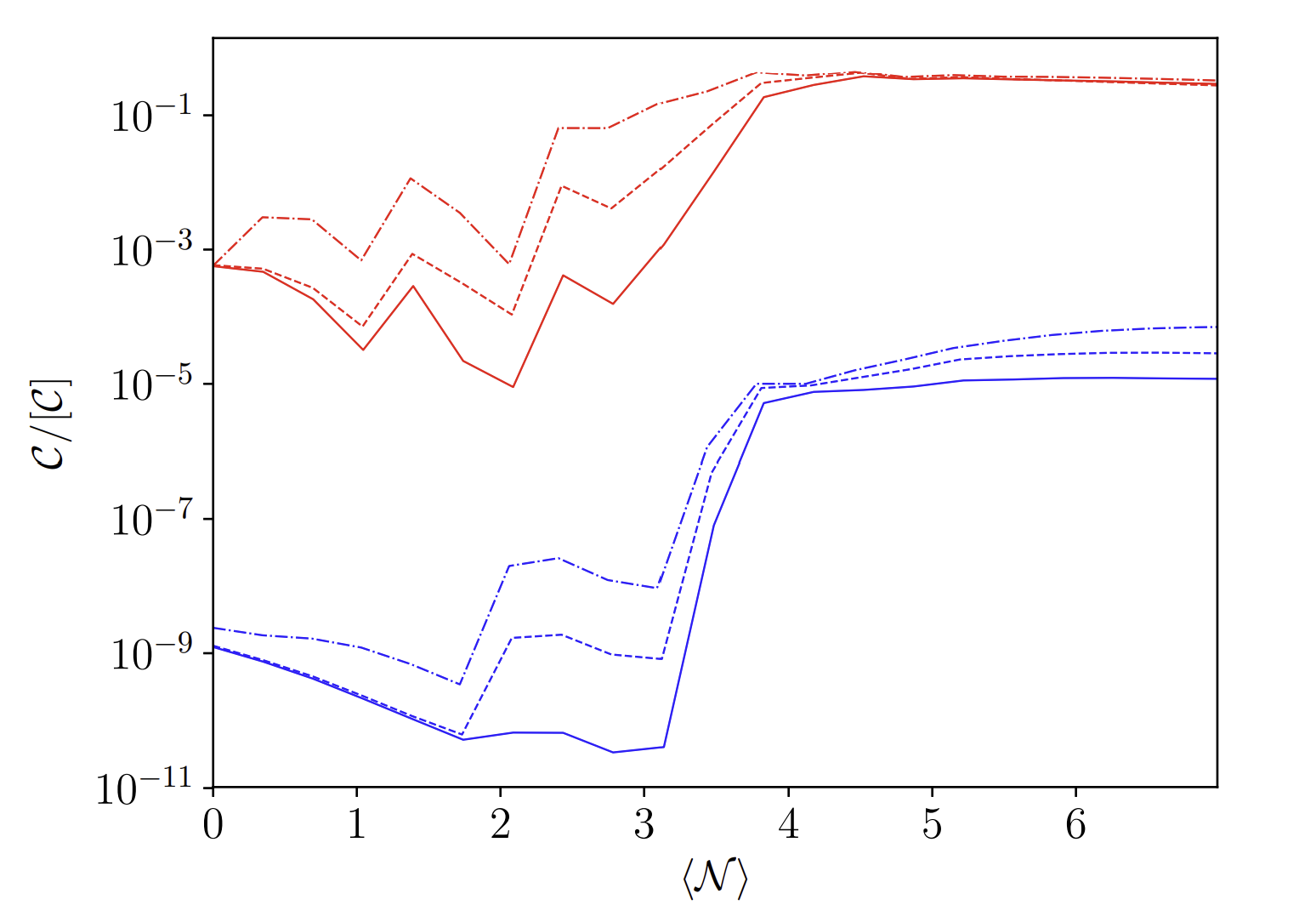}
\caption{ \justifying Strong resonance convergence test for the mean hamiltonian (blue) and momentum (red) relative constraints  ($\langle|{\cal C}|/[{\cal C}]\rangle$, ${\cal C }= {\cal H}, {\cal M}$) across multiple runs with $N=64$ (dash-dotted), $N=128$ (dashed), and $N = 256$ (solid).  With the same range of scales initially, the perturbative phase gets smaller constraints with better resolution, as opposed to the non-pertubative last time steps of the momentum constraint.}
\label{fig:mono_convergence}
\end{figure}
\begin{figure*}
    \centering
    \begin{subfigure}{\textwidth}
\includegraphics[width=0.85\linewidth]{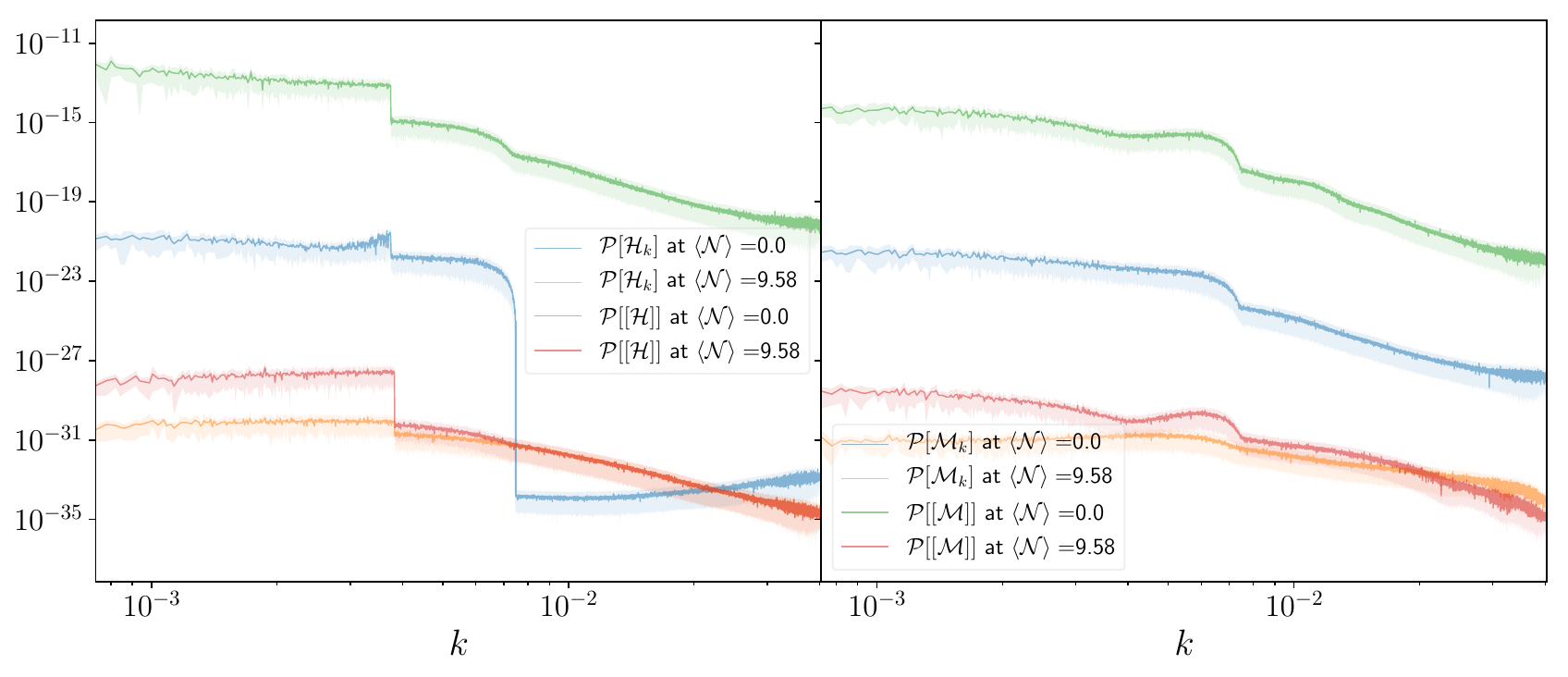}
\caption{Inflection inflation}
    \end{subfigure}
    \begin{subfigure}{\textwidth}
\includegraphics[width=0.85\linewidth]{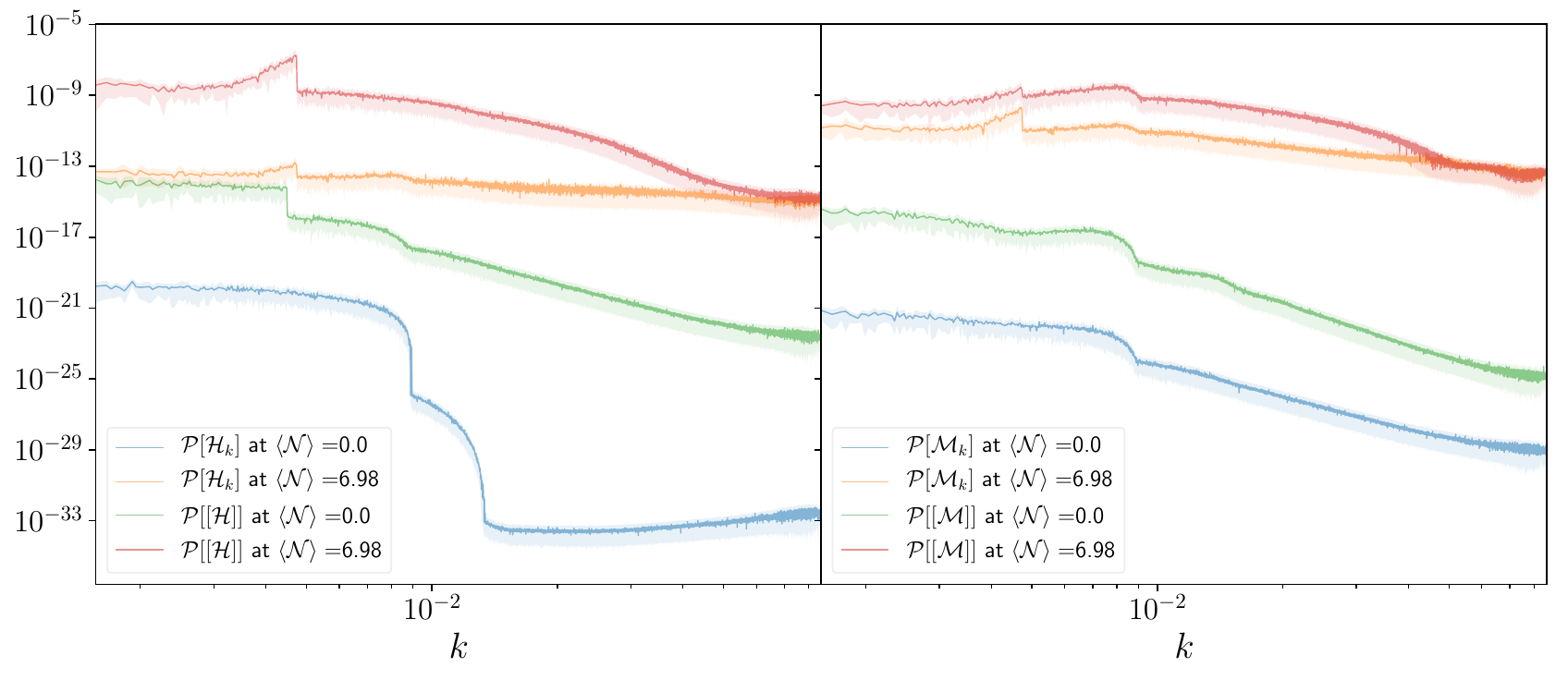}
\caption{Strong resonance }
    \end{subfigure}
\caption{\justifying Initial and final spectra of the hamiltonian (left) and momentum (right) constraint violations (blue and orange) and absolute constraint magnitudes (green and red) for our non-perturbative examples. {The constraints are well satisfied if the orange line lies below the red.} The spectral input was cut at $k_{max}^{\cal R} = 4\times 10^{-3}$ and $k_{max}^{\cal R} = 4.5\times 10^{-3}$ respectively, visible through spectral drops. $M_{Pl}=100$ units.}
\label{fig:specCons}
\end{figure*}

The two non-perturbative examples studied in \ref{subsec:inflection} and \ref{subsec:strongRes} should be of great interest not only to cosmologists but also to the numerical relativity community. In one case the field slows down to a barely resolvable speed (ultra slow-roll reached during inflection) and, in the other case, domain walls form during the strong resonance of monodromy inflation.   These nonlinear multi-scale effects provided challenging tests of the present implementation of our methods, at the same time demonstrating their robustness, while defining some outer limits.

The degree of accuracy of the numerical evolution can be observed from the full Hamiltonian ${\cal H}/[{\cal H}]$ and momentum ${\cal M}/[{\cal M}]$ constraints shown in Figure \ref{fig:NPconstraints}.  Both ratios  ${\cal H}/[{\cal H}]$ and  ${\cal M}/[{\cal M}]$ should remain much smaller than $1$ for correctness, whether the simulation is perturbative or not. Both scenarios show good constraint preservation at the initial linear level during at least the first half of the simulations, that is, for over 7 e-folds and 3 e-folds respectively for ultra slow-roll and monodromy inflation.  {However, a problem develops} in the momentum constraint, while the hamiltonian constraint (upper panels of Figures \ref{fig:NPconstraintsa} and \ref{fig:NPconstraintsb}) remains satisfied within a  margin $\lesssim 10^{-3}\ll 1$ at all times and {grid points}. One could think that increasing the resolution would help reduce the relative momentum constraints to values smaller than $1$ {but, as might be guessed from the convergence test provided in Figure \ref{fig:mono_convergence}, increasing the resolution with the same input does not {however} improve the momentum constraint at late non-perturbative times}. 

It is useful to understand which scales {are responsible for such constraint violation} using the Fourier transform of the constraints. This is highlighted in Figure \ref{fig:specCons}: with initial UV mode cutoffs at $k = 4\times 10^{-3}$ and $ 4.5 \times 10^{-3}$ for inflection and resonance cases respectively, it is very clear that the constraints and their absolute magnitudes only overlap at late times for modes on scales smaller than the ones included in the initial spectrum, which are scales where gradient computations fail. This momentum constraint breakdown in the axion monodromy case is primarily due to numerical resolution effects after the formation of domain walls, which is a well known problem for the numerical  evolution of topological defects in an expanding universe. This is evidenced by the fact that there is only localised breakdown of the constraints {along the domain walls (about 15\% of the grid points)}, rather than a generic problem throughout the simulation (e.g.\ it is not due to non-perturbative effects from strong gravity in this case). When these later emerging, constraint violating UV scales (we choose conservatively $k>10^{-2}$) are filtered out from the Fourier transform of the various fields, the deviations from the FLRW background shown in fig.~\ref{fig:meanmono} are not affected. Furthermore, we note that if the simulation of the monodromy potential is stopped at $3.8$ efolds, just as the bimodal distribution starts forming and while the majority of points still respect the momentum constraint, the formation of bubbles of trapped fields is still observed. Hence, the non-perturbative effects of the monodromy potential observed in our simulations are robust against brute force removal of constraint violating Fourier modes.\footnote{A more sophisticated dynamical investigation of such mode filtering is beyond the scope of this paper.}

Accurate numerical resolution of the emerging domain walls on which the momentum constraint currently fails in our simulations is a challenging endeavour: such domain walls have a fixed physical width $\delta_{\rm dw} $ that is determined by the inverse mass associated with symmetry breaking in the potential, $\delta_{\rm dw} \sim m^{-1}_{\rm dw}$. On the other hand, the physical length scale related with the comoving grid scale $\triangle x$ expands rapidly as $\triangle r(t)  = a(t) \triangle x$ (especially during inflation).  So this soon exceeds the defect width $\triangle r(t)  > \delta_{\rm dw}$ meaning that there is insufficient numerical resolution to evolve the domain walls accurately.  This is a problem for which there have been a number of alternative solutions developed\footnote{One frequently used approach is the `fat defect' algorithm, pioneered to study axion domain walls in an expanding universe \cite{Press:1989yh}; this essentially makes the mass dynamical $m_{\rm dw}\propto a(t)$, so the domain walls have a fixed comoving width. However, despite the simplicity of this approach, it modifies the underlying physics of the defect network, notably its gravitational effects.}, the best of which appears to be the use of {adaptive} mesh refinement (AMR) (see, for example, ref.~\cite{Drew:2019mzc}).  We note that \textsc{STOIIC} is compatible with GRTeclyn, which is built on the AMReX libraries, so it has AMR as a core capability. However, we have not yet deployed this to push further forward and maximise dynamical range, because complex topological defect evolution is not the primary focus of this paper. Nevertheless, we highlight this additional AMR capability which should also improve the evolution of any large perturbations, also in the ultra slow-roll scenario.  In any case, despite the emergence of numerical inaccuracies late in the evolution, we note that they do not cause a more serious numerical breakdown beyond the momentum constraint because the small-scale perturbations become superhorizon and their evolution essentially freezes. 

{We would like to comment further on the importance of independently monitoring the accuracy of constraint evolution during cosmological simulations. Satisfying the momentum constraint often proves to offer an extra layer of challenge in numerical relativity due to the spatial gradients involved. To our knowledge, this has not often been explicitly diagnosed in previous inhomogeneous inflationary simulations, sometimes unmentioned, sometimes relying on an initial trivial satisfaction \cite{clough_robustness_2017} or the long wavelength approximation \cite{Caravano2024}. However, it is known that the momentum constraint is a key connector of the Hubble-sized universe patches \cite{prokopec_ensuremathdeltan_2021} and should not be overlooked on those scales.   In our own simulations, we have implemented a self-consistent methodology for the initial conditions to ensure that the constraints are satisfied at linear order (i.e.\ effectively with violations only appearing at second-order). In the present simulation we have stress-tested this approach by setting the whole grid initially inside one Hubble volume $L^{\circledast} \lesssim (H^{\circledast})^{-1}$.  This UV depth of our modes is especially large for axion monodromy where the most subhorizon simulations are necessary to observe large resonances. In general, Bunch-Davies fluctuations deep inside the horizon rise in amplitude and become less and less perturbative (leaving aside the question of whether it is a good approximation to neglect subhorizon quantum effects (see, e.g., ref.~\cite{Launay2024bis})). Most of our simulations, such as the previous quadratic case, satisfy the constraints accurately and will thus allow us to look more closely into these issues. The two non-perturbative examples presented here are certainly some of the most extreme cases, where the satisfaction of the momentum constraint does not improve over time, but at least the regime of validity of the resulting predictions are well defined. }

Another point of debate, mentioned in \cite{aurrekoetxea24}, is that of gauge choice. In principle, the geodesic gauge will end up ill-posed when gravitational collapse becomes possible, because of the crossing of free-falling observers on the grid \cite{baumgarte_shapiro_2010}. In this case, the spatial coordinates can converge, intersect, and so leave a coordinates singularity that prevents further simulation evolution. This is not as visible in inflationary simulations as in strong gravity scenarios, such as black hole simulations, but it is there, especially if we continue beyond reheating.  There is some evidence of this in the gradual growth of the spatial metric determinant contrast: the ratio $\delta X / X_b $ increases slowly with time according to eq.~\eqref{eq:ADMgaugeDecomp} and Appendix \ref{app:dicBSSN}, notably because of the growing $E$ and $\chi$ contributions. This is shown in Figure \ref{fig:Xcontrasts}, even for the simple slow-rolling phases, while staying fully perturbative for the whole duration of our simulations, so we conclude that using the geodesic gauge is not problematic. In future, we plan to compare this evolution with another successful gauge, adapted from the integrated moving puncture gauge to the case of inhomogeneous inflation \cite{giblin_jr_preheating_2019, Kou21, aurrekoetxea_oscillon_2023}.

Our study of higher-order interactions and correlation functions is reserved for future work. A consequent study will indeed be needed to prove that these simulations can reach very high orders of perturbation theory that have not been achieved analytically (at least quantitatively). The manner in which the inflationary initial data is input and windowed might significantly impact the predictions. In our present simulations, we are essentially setting a cutoff, and while this should not impact the dynamics, it does mean that modes are not treated equally as they do not enter the horizon at the same stage of their interactions \cite{Launay2024bis}. Stochastic inflation might prove a better answer because of the time-dependent cutoff, improving the accuracy of the constraints; a forthcoming study in full numerical relativity will describe the implementation of the stochastic approach described in \cite{Launay24}.

%

\cleardoublepage
\section{Conclusions and future challenges \label{sec:sec5}}
In this work, we pioneered the full evolution of the Bunch-Davies vacuum initial conditions for inflation in numerical General Relativity. We first solved the generic initial conditions problem for inhomogeneous inflation at linear order in scalar perturbations (including the metric) for a common gauge: the geodesic (or synchronous) gauge. We emphasize that the method itself is not restricted to this particular gauge choice and could be adapted to other choices also. The rest of the paper then adapted Numerical Relativity techniques to cosmology, building up from previous works \cite{aurrekoetxea24}. Three examples were studied, ranging in the level of emerging non-linearity: a very perturbative and linear case of quadratic inflation, a potential with an inflection point and a strongly non-perturbative and non-linear resonance monodromy model. All evolutions including the most exotic ones satisfy the constraints for our scales of interest. We proved that looking at non-perturbative scenarios is possible and showcases new phenomena such as highly inhomogeneous geometries and fields one can find from a monodromy model. We also commented on the general need to use observer-independent, non-linear variables (like $\mathcal{R}_i$) in order to extract measures of the fluctuations from our numerical boxes.

This work demonstrates that it is possible to sidestep perturbation theory in making (numerical) predictions for post-inflation states involving fully General Relativistic dynamics. In future work, we aim to make more refined predictions with bigger simulations and better estimators for the fluctuations and corresponding observables. Important improvements might further involve setting up the initial conditions beyond first order by combining our approach with a numerical solver like \cite{aurrekoetxea_cttk_2022,GRTresna}, studying the evolution in different coordinates \cite{giblin_jr_preheating_2019,Kou21,aurrekoetxea_oscillon_2023}, incorporating recent work to include gravitons \cite{Florio2024}, implementing multiple fields, etc. In particular, we hope to focus on model-specific studies, for instance, to gain more accurate insights on ultra slow-roll and primordial black hole production. 

Most importantly, as advocated previously in \cite{Launay24,Launay2024bis}, the key extension of this work is the incorporation of stochastic inflation. In this framework, the modes do not all enter the simulation at the same initial time but at a given elapsed time relative to their Hubble horizon crossing, thus better accounting for quantum diffusion. As recently announced in \cite{Launay24}, we will soon report on the possibility of running stochastic inflation in full numerical relativity and thus allowing comparison with the present work that initializes the evolution from a single time-slice, as in a standard initial value numerical set up, similarly to lattice cosmology. 
Hopefully, the numerical implementation of the prescription described in \cite{Launay2024bis} will pave the way for non-pertubative precision predictions from inflation, for instance on primordial non-gaussianities or non-linear sub-CMB scales, where constraints on gravitational waves and primordial black holes could be enhanced.

\newpage
\acknowledgments
Y.L. would like to thank several individuals for discussions related to this project. For their advice, expertise and excellent work on developing core software for the GRTL collaboration, warmest thanks go to Juliana Kwan and Miren Radia, supported by the Intel oneAPI COE and STFC DiRAC. Other members of the collaboration are to be thanked, such as Amelia Drew, Matthew Elley, Cristian Joana, Panos Giannadakis, and Eugene Lim, and in particular Ericka Florio and Josu Aurretexoa for their regular help and opinions. Thanks are also to be given for many useful conversations to other colleagues: Thomas Colas, Ciaran McCulloch, David Baker, Enrico Pajer in DAMTP, but also Swagat Mishra, Ian Moss, Angelo Caravano. 
 Y.L. is supported by the STFC DiS-CDT scheme and the Kavli Institute for Cosmology, Cambridge. E.P.S.S. acknowledges funding from STFC Consolidated Grant No. ST/P000673/1. Computational resources were supported by DiRAC resources and a grant from G-research.
{The authors all thank the anonymous referee for their advice, which helped improving the manuscript.}

\newpage
\appendix

\section{Linearised BSSN dictionary \label{app:dicBSSN}}
Using ADM and scalar SVT notations introduced in sections \ref{subsec:initValprob} and \ref{subsec:Inhom-Constr}, we derive
\begin{itemize}
    \item[\ding{118}]${X} = a^{-2} [1+2\Phi -\frac{2}{3}\nabla^2 E]. $\\
            \textit{proof.}  $^{(3)}g = det(^{(3)}g_{ij})= a^6 [1-6\Phi +2(E_{,11}+E_{,22}+E_{,33})] = a^6 [1-6\Phi +2\nabla^2 E]$ \\
            Hence $[^{(3)}g]^{-1/3} = a^{-2} [1+2\Phi -\frac{2}{3}\nabla^2 E].$
            
     \item[\ding{118}]${\tilde{\gamma}_{ij}} = \delta_{ij}+2(\partial_i\partial_j-\frac{1}{3}\delta_{ij}\nabla^2)E,$ and ${\tilde{\gamma}^{ij}} = \delta^{ij}-2\delta^{im}\delta^{jn}(\partial_m\partial_n-\frac{1}{3}\delta_{mn}\nabla^2)E.$ \\
      \textit{proof.}  $X{^{(3)}g_{ij}} = a^{-2} [1+2\Phi -\frac{2}{3}\nabla^2 E] a^2 [(1-2\Phi)\delta_{ij}+\partial_{ij}E] = \delta_{ij}+2E_{,ij}-\frac{2}{3}\delta_{ij}\nabla^2E$. The inverse metric is obtained at first order with $\tilde{\gamma}^{ik}\tilde{\gamma}_{kj} = {\delta}{^i{}_j}$.
      
      \item[\ding{118}]${K} =  -3H+\kappa.$ \\
      Using the definition of $K$ from eq. \ref{eq:eqK}: $\kappa = 3(\frac{\dot{\Phi}}{\overline{N}}+H\Psi)-\Delta \chi$ and $\chi = -\frac{a^2}{\overline{N}}(B-\dot{E})$.
      \item[\ding{118}]${\tilde{A}_{ij}} =  \frac{1}{\overline{N}}(\partial_i\partial_j-\frac{1}{3}\delta_{ij}\nabla^2)(B-\dot{E}).$ \\
      \textit{proof.} Similarly $K_{ij}^{TF} = -(\partial_i\partial_j-\frac{1}{3}\delta_{ij}\nabla^2)\chi$  \\
      so  $X K_{ij}^{TF} =a^{-2} [1+2\Phi -\frac{2}{3}\nabla^2 E]  \frac{a^2}{\overline{N}}(\partial_i\partial_j-\frac{1}{3}\delta_{ij}\nabla^2)(B-\dot{E}) = \frac{1}{\overline{N}}(\partial_i\partial_j-\frac{1}{3}\delta_{ij}\nabla^2)(B-\dot{E}).$
    \item[\ding{118}]${\alpha} = \overline{N}(1+\Psi). $ 
     \item[\ding{118}]${\beta_{i}} = a^2B_{,i}.$

    \item[\ding{118}]${\tilde{\Gamma}^i_{jk}} = \delta^{il} E_{,lkj} - \frac{1}{3}\nabla^2[E_{,j}{\delta}{^i_k}+E_{,k}{\delta}{^i_j}-E_{,l}\delta^{il}\delta_{jk}].$\\
    \textit{proof.} $\tilde{\Gamma}^i_{jk} = \frac{1}{2}[ \delta^{ij}-2\delta^{im}\delta^{jn}(\partial_m\partial_n-\frac{1}{3}\delta_{mn}\nabla^2)E][2\partial_j(\partial_l\partial_k-\frac{1}{3}\delta_{lk}\nabla^2)+2\partial_k(\partial_j\partial_l-\frac{1}{3}\delta_{jl}\nabla^2)-2\partial_l(\partial_j\partial_k-\frac{1}{3}\delta_{jk}\nabla^2)]E 
    = \delta^{ij}[\partial_j(\partial_l\partial_k-\frac{1}{3}\delta_{lk}\nabla^2)+\partial_k(\partial_j\partial_l-\frac{1}{3}\delta_{jl}\nabla^2)-\partial_l(\partial_j\partial_k-\frac{1}{3}\delta_{jk}\nabla^2)]E
    = E_{,lkj}\delta^{il}-\frac{1}{3}\nabla^2[E_{,j}{\delta}{^i_k}+E_{,k}{\delta}{^i_j}-E_{,l}\delta^{il}\delta_{jk}]+o(E).$
    \item[\ding{118}]${\tilde{\Gamma}^i} = \frac{4}{3}\delta^{ij}\nabla^2E_{,j}. $\\
       \textit{proof.} Computing $\tilde{\gamma}^{jk} \tilde{\Gamma}^i_{jk} = \delta^{jk} \tilde{\Gamma}^i_{jk},$ as previous expression is a first order.
    \item[\ding{118}]${R_{ij}} = (\nabla^2\delta_{ij} + \partial_i\partial_j)\Phi,$ and  ${R^{TF}_{ij}} = (\partial_i\partial_j-\frac{1}{3}\delta_{ij}\nabla^2)\Phi.$\\
    \textit{proof.} One can show that ${^{(3)}{R}{^i{}_j}} = \Delta \Phi \delta^k{}_j+\Phi^{,k}{}_{,j}$ and lower the index.
\end{itemize}

\section{Cosmological extractions \label{app:extract}}
\textsc{VIZIR}'s postprocessing makes use of diverse relations to extract diagnostics and observables. Here are the most important relations, $\langle \rangle_{\scriptscriptstyle \square}$ denoting a box averaging
\begin{equation}
    \left\{\begin{aligned}\langle {\cal N} \rangle_{\scriptscriptstyle \square} & = -\frac{1}{2}\ln {\langle X \rangle_{\scriptscriptstyle \square}}, \\
    \Phi_k & = \frac{1}{2}(a^2 X_k-1)-\frac{1}{48}\sum_{i,j=1}^3 \frac{k_ik_j}{k^2}(\tilde{\gamma}_{ij}-1),\\
    E_k & = -\frac{1}{16}\sum_{i,j=1}^3 \frac{k_ik_j}{k^4}(\tilde{\gamma}_{ij}-1),\\
   {\cal R} &= \Phi -\frac{\langle K \rangle_{\scriptscriptstyle \square}}{\langle \alpha \rangle_{\scriptscriptstyle \square}\langle \Pi \rangle_{\scriptscriptstyle \square}}(\phi-\langle \phi\rangle_{\scriptscriptstyle \square}).
   \end{aligned}\right .
\end{equation}

\section{Randomness with a power spectrum \label{app:random}}
As a reminder, the Fourier transform ${\cal F}$ and its inverse are defined with a backward normalisation in QFT. 

We define RFFTN (resp. IRFFTN) as a discrete Fourier transform (resp. its inverse) on a positive 3D hemisphere with no normalisations ($RFFTN\{IRFFTN\}=N^3$).

We want to draw a random 3D process with power spectrum $|{f}_{\vec{k}}|^2$ defined from QFT as
 \begin{equation}
  \hat{f}(x) = {\cal F}^{-1}\{ {f}_{\vec{k}}\hat{a}_{\vec{k}}\} + h.c.,
\label{eq:ModesExpApp}
 \end{equation}
There are a few subtleties one needs to account for.
\begin{itemize}
\item[\ding{118}] Assuming that units are such that $[\hat{f}(x)] = M^0$, we understand from the consequently unitless two-points function
\begin{equation}
\langle \hat{f}(x)^2\rangle = {\cal F}^{-1}\{ |{f}_{\vec{k}}|^2 \},
\end{equation}
that anihilation and creation operators hide units and that ${f}_{\vec{k}}$ is not a simple Fourier density ($[{f}_{\vec{k}}] \neq M^{-3}$) but such that $[{f}_{\vec{k}}] = M^{-3/2}$.

\item[\ding{118}] Building a real stochastic Fourier expansion doesn't require the addition of the hermitian conjugate\footnote{If we were doing so, the stochastic approach would require a factor $\sqrt{2}$ due to the commutativity of the latter and not of  QFT.} because we are dealing with numbers (we can find $\boldsymbol{\alpha}$ such that $\boldsymbol{\alpha}_{-\vec{k}}= \boldsymbol{\alpha}^*_{\vec{k}}$ while $\hat{a}_{-\vec{k}}\neq \hat{a}^\dagger_{\vec{k}}$) 
\begin{equation}
  \boldsymbol{f}(x) = {\cal F}^{-1}\{ \tilde{f}_{\vec{k}}\boldsymbol{\alpha}_{\vec{k}}\} 
\end{equation}
where $\tilde{f}_{\vec{k}} = f_k$ (resp. $f_k^*$) if $\vec{k}$ is in the $k_z>0$ hemisphere (resp. $k_z<0$ hemisphere). This is essentially saying that we impose a hermitian symmetry as (I)RFFTN does.
We can compute the coincident two-point function
\begin{equation}
    \left\{
    \begin{aligned}
        \langle \boldsymbol{f}(x)^2 \rangle & = \int\frac{d^3\Vec{k_1}}{{(2\pi)}^3}\frac{d^3\Vec{k_2}}{{(2\pi)}^3}\tilde{f}_{\vec{k}_1}\tilde{f}_{\vec{k}_2}\langle \boldsymbol{\alpha}_{\vec{k}_1}\boldsymbol{\alpha}_{\vec{k_2}}\rangle\\
        &  = \int\frac{d^3\Vec{k_1}}{{(2\pi)}^3}\frac{d^3\Vec{k_2}}{{(2\pi)}^3}\tilde{f}_{\vec{k}_1}\tilde{f}_{-\vec{k}_2}\langle \boldsymbol{\alpha}_{\vec{k}_1}\boldsymbol{\alpha}_{\vec{k_2}}^*\rangle \\
 & = \frac{C^2}{(2\pi)^3}\int\frac{d^3\Vec{k}} {{(2\pi})^3}|f_k|^2 = \frac{C^2}{(2\pi)^3}\langle \hat{f}(x)^2\rangle, 
    \end{aligned}\right .
\end{equation}
where the integral in $\vec{k}_2$ has been reversed and the hernitian symmetry of $\boldsymbol{\vec{k}}$ assumed, before applying eq. \eqref{eq:stocomm}.
 The right normalisation is the one that matches the stochastic and quantum correlations functions, thus $C = (2\pi)^{3/2}$.

\item[\ding{118}] To generate the $\boldsymbol{\alpha}_{\vec{k}}s$, it is possible to use
\begin{equation}
     \boldsymbol{\alpha}_{\vec{k}}/C = \boldsymbol{s}_{\vec{k}} = \frac{1}{\sqrt{N}^3}\text{RFFTN}\{\Delta(\vec{x})\}(\vec{k}), 
\end{equation}
when $N$ is big enough and taking each $\Delta(\vec{x})$ as independent random normal variables on the grid. This will indeed imply the statistics of eq. \eqref{eq:stocomm} and their hernitian symmetry.
\end{itemize}
This leads to the final strategy:
\begin{equation}
\left\{
\begin{aligned}
   \boldsymbol{f}(x) 
 & =  \displaystyle \left(\frac{dk}{2\pi}\right)^{3}\times\text{IRFFTN}\left\{ \left(\frac{dk}{2\pi}\right)^{-\frac{3}{2}}f_{\vec{k}}\times (2\pi)^{3/2}\boldsymbol{s}_{\vec{k}}\right\} \\
 & = \displaystyle\left(\frac{dk}{\sqrt{2\pi}}\right)^{\frac{3}{2}}\times\text{IRFFTN}\left\{ f_{\vec{k}}\boldsymbol{s}_{\vec{k}}\right\}.
 \end{aligned}\right .
\end{equation}

\begin{figure}
    \centering
        \begin{subfigure}{0.5\textwidth}
\includegraphics[width=0.85\linewidth]{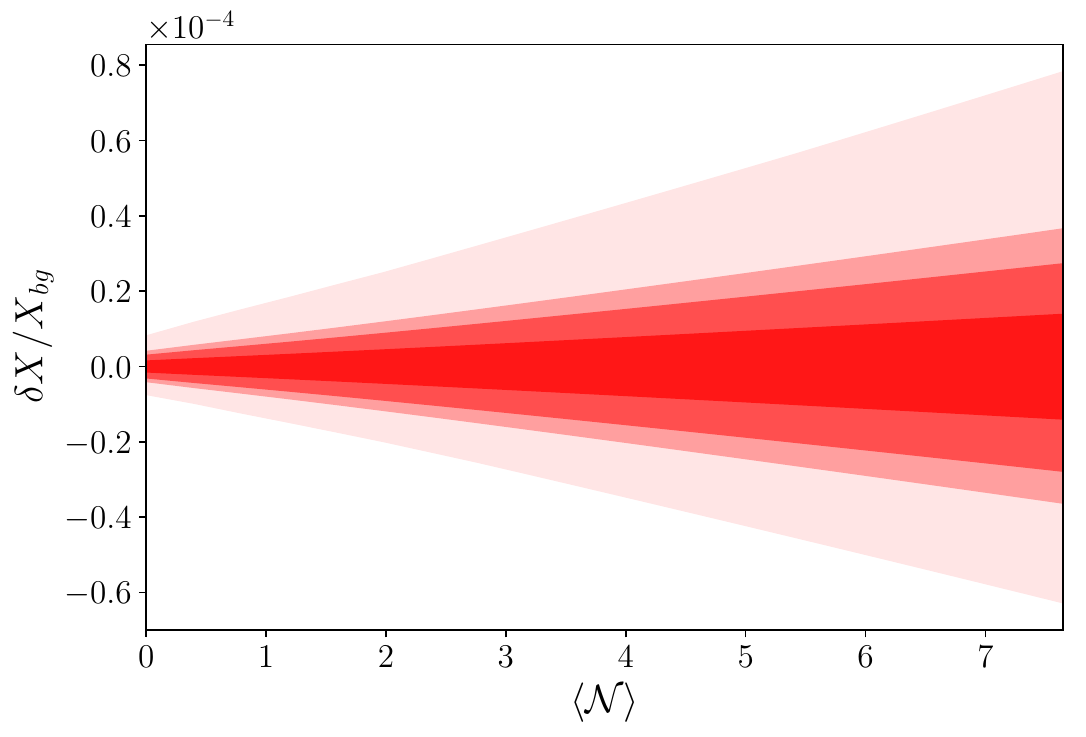}
\caption{Quadratic}
    \end{subfigure}
    \begin{subfigure}{0.5\textwidth}
\includegraphics[width=0.85\linewidth]{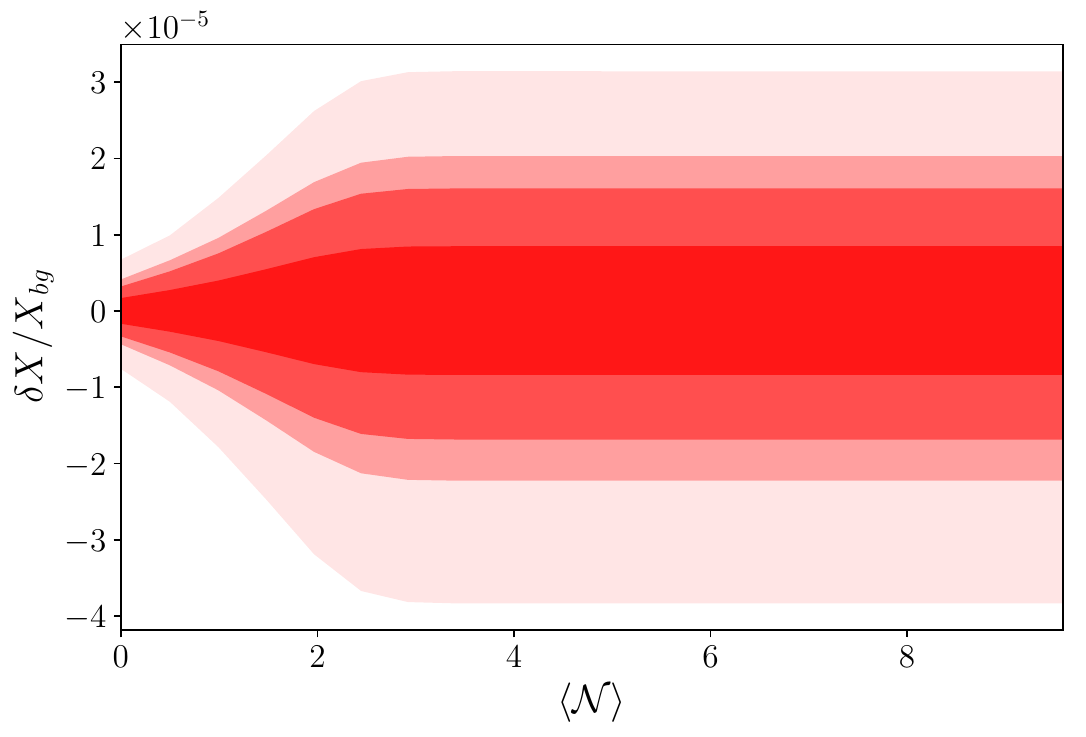}
\caption{Inflection}
    \end{subfigure}
    \begin{subfigure}{0.5\textwidth}
\includegraphics[width=0.85\linewidth]{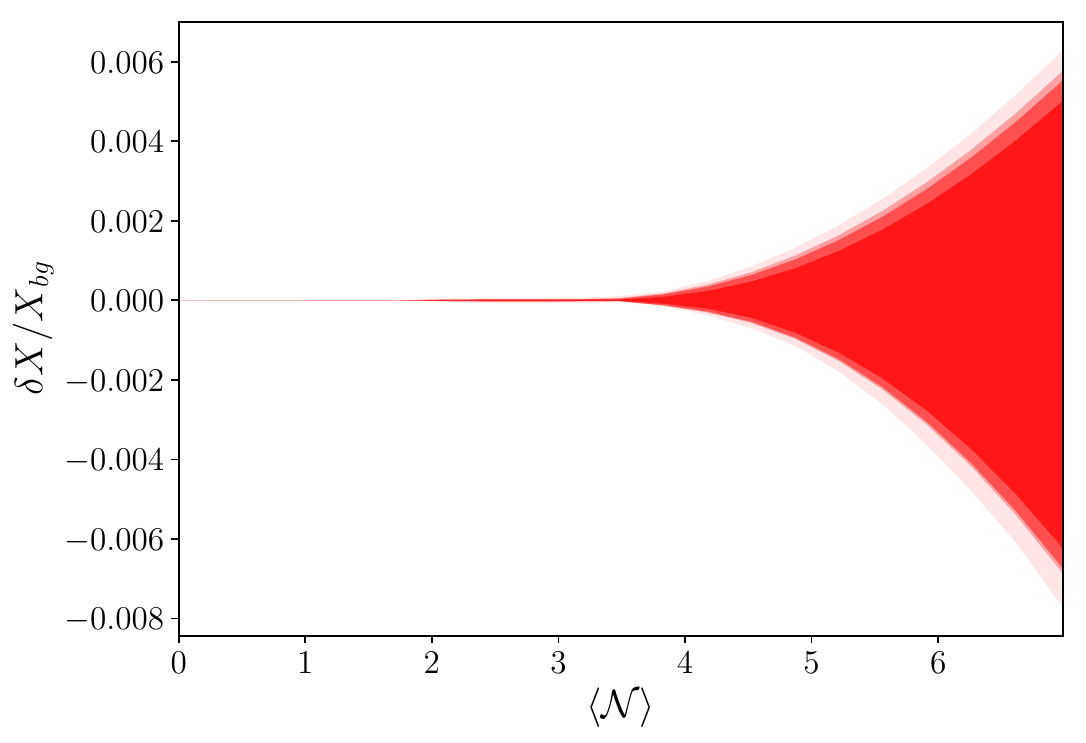}
\caption{Strong resonance}
    \end{subfigure}
    \caption{\justifying $68$, $95$, $99$ and $100$ percentile contours for the contrast of $X$.}
    \label{fig:Xcontrasts}
\end{figure}

\section{Non-perturbative comoving curvature \label{app:RNL}}
The non-perturbative comoving curvature is defined as \cite{rigopoulos_non-linear_2005, langlois_evolution_2005}
\begin{equation}
    {\cal R}_i = -\frac{1}{6}\partial_i\ln |\gamma|+\frac{1}{6}\frac{\partial_0\ln |\gamma|}{\partial_0 \phi} \partial_i \phi,
\end{equation}
and constitutes a gauge-invariant quantity.

In particular, its linearisation writes 
\begin{equation}
\left\{\begin{aligned}
     {\cal R}_i^{(1)} & = {\cal R}_{,i} -\Phi(t_0)_{,i}-\frac{1}{3}\nabla^2 [E-E(t_0)]_{,i}+\frac{1}{3}\int_{t_0}\nabla^2B_{,i}\\
     & = {\cal R}_{,i}-\Phi_{B}(t_0)_{,i}-\frac{1}{3}\nabla^2E_{,i}.
\end{aligned}\right .
   \label{eq:RNLlindem}
\end{equation}
Note that the two last terms are usually very suppressed. This expression explains why defining ${\cal R}^{NL}_k = i\sum_a \frac{k_a}{k^2}({\cal R}_a+\Phi_B(t_0)-\frac{1}{3}\nabla^2E)$ is a good way to compare ${\cal R}_i$ and ${\cal R}$.

\section{BSSN equations \label{app:BSSN}}
\begin{equation}
\left\{
\begin{aligned}
 \partial_t X -\frac{2}{3} X \alpha K+\frac{2}{3} X \partial_k \beta^k-\beta^k \partial_k X = 0,&\\
 \partial_t \tilde{\gamma}_{i j}+2 \alpha \tilde{A}_{i j}-\tilde{\gamma}_{i k} \partial_j \beta^k-\tilde{\gamma}_{j k} \partial_i \beta^k &\\
 \quad+\frac{2}{3} \tilde{\gamma}_{i j} \partial_k \beta^k-\beta^k \partial_k \tilde{\gamma}_{i j} =0, &\\
  \partial_t K+\gamma^{i j} D_i D_j \alpha-\alpha\left(\tilde{A}_{i j} A^{i j}+\frac{1}{3} K^2\right) &\\
-\beta^i \partial_i K-4 \pi \alpha(\rho+S) = 0, &\\
 \partial_t \tilde{A}_{i j}-X\left[-D_i D_j \alpha+\alpha\left(R_{i j}- M_{Pl}^{-2}\alpha S_{i j}\right)\right]^{{TF}} &\\
-\alpha\left(K \tilde{A}_{i j}-2 \tilde{A}_{i l} \tilde{A}_j^l\right)
-\tilde{A}_{i k} \partial_j \beta^k-\tilde{A}_{j k} \partial_i \beta^k &\\
+\frac{2}{3} \tilde{A}_{i j} \partial_k \beta^k-\beta^k \partial_k \tilde{A}_{i j} = 0, &\\
 \partial_t \tilde{\Gamma}^i-2 \alpha\left(\tilde{\Gamma}_{j k}^i A^{j k}-\frac{2}{3} \tilde{\gamma}^{i j} \partial_j K-\frac{3}{2} A^{i j} \frac{\partial_j X}{X}\right) &\\
+2 A^{i j} \partial_j \alpha-\beta^k \partial_k \tilde{\Gamma}^i 
-\tilde{\gamma}^{j k} \partial_j \partial_k \beta^i-\frac{1}{3} \tilde{\gamma}^{i j} \partial_j \partial_k \beta^k &\\
-\frac{2}{3} \tilde{\Gamma}^i \partial_k \beta^k+\tilde{\Gamma}^k \partial_k \beta^i+ 2 M_{Pl}^{-2}\alpha \tilde{\gamma}^{i j} S_j = 0, &\\
 \partial_t \phi-\alpha \Pi-\beta^i \partial_i \phi = 0, &\\
 \partial_t \Pi-\beta^i \partial_i \Pi-\alpha \partial_i \partial^i \phi-\partial_i \phi \partial^i \alpha &\\
-\alpha\left(K \Pi-\gamma^{i j} \Gamma_{i j}^k \partial_k \phi-\frac{d V}{d \phi}\right)= 0,  &\\
         \mathcal{H}=R+\frac{2}{3}K^2-\tilde{A}_{i j} A^{i j}- 2 M_{Pl}^{-2}\rho = 0, & \\
        \mathcal{M}_i=\tilde{\gamma}^{k l}\left(\partial_k \tilde{A}_{l i}-2 \tilde{\Gamma}_{l(i}^m \tilde{A}_{k) m}-3 \tilde{A}_{i k} \frac{\partial_l X}{2 X}\right) & \\
        -\frac{2}{3} \partial_i K- M_{Pl}^{-2}S_i =0. &
    \end{aligned}\right .
        \label{eq:BSSN}
\end{equation}

\section{Warning on looking to deep in the UV \label{app:UVdep}}

\begin{figure}
    \centering
\includegraphics[width=0.95\linewidth]{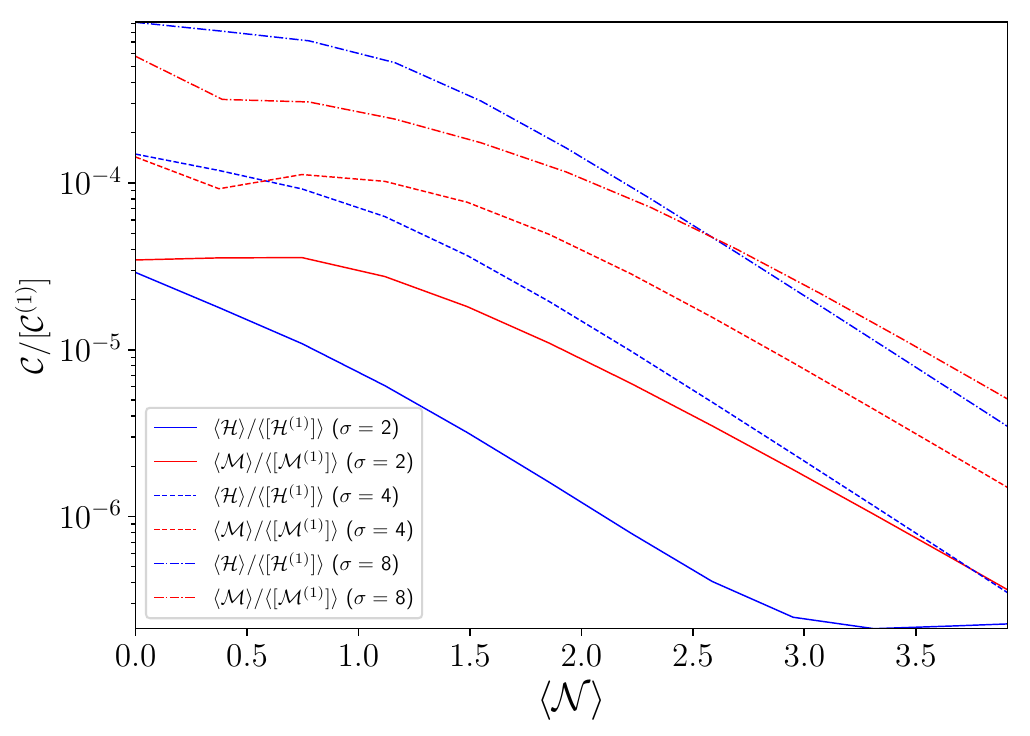}
    \caption{\justifying Mean constraint violation to mean first order magnitude ratio  ($\langle|{\cal C}|/[{\cal C}^{(1)}]\rangle$, ${\cal C }= {\cal H}, {\cal M}$) for quadratic inflation with different UV cutoffs. For $\sigma = 1$ we set $L = 128/H^{\circledast}$ and adapt $L$ when changing $\sigma$ to keep their product a constant.\footnote{ Note that the hamiltonian $\sigma=2$ final ratio flattens because of the constraint violation reaching numerical accuracy (smaller $\sigma$ means smaller perturbations and so violation).}}
    \label{fig:sigdep}
\end{figure}

We demonstrate here the breakdown of our methods when including modes that are too subHubble. When varying $\sigma$, one can study the scaling of the simulation independently from the resolution by rescaling the resolution accordingly, for instance by rescaling $L$. Thanks to this, gradients remain perfectly accurate and cannot violate the constraints with amplitude greater than second order in perturbation theory. Figure \ref{fig:sigdep} showcases such an experiment with the case of quadratic inflation. It shows the ratio of the constraints violation and the typical amplitude of a first-order violation of the constraints $|{\cal H}|/[{\cal H}^{(1)}]$ and $|{\cal M}|/[{\cal M}]$. Given that our method provides violation at second order in perturbation theory, this ratio is a direct comparison of the second and first orders.

Cutting more of the UV makes the contrasts in the box decrease, which explains why the constraints do too. Inversely, the contrasts and so the constraints increase when including more and more UV modes, which diverge in this limit. In particular, the momentum's contrast in the geodesic gauge has usually been the first to break perturbation theory in the UV in this work.

When increasing $\sigma$, one has to stop before exiting the perturbative regime. Indeed, the second-order perturbations increase in amplitude, and the further you move in the UV the bigger the effect. Our constraints are still satisfied to first order but with a smaller gap between each order of perturbation theory. As this gap decreases by including more UV modes, satisfying the constraints at linear order as we do here is no longer sufficient and another approach must be used. However, this does not seem to be the real problem as there is no certainty on the physicality of those deep subHubble modes. While some of us argue that a regularization should be performed, for instance using the adiabatic method \cite{Durrer09, pla23}, we will simply state that plugging those diverging modes in Einstein's equations and so evolving them in NR would probably lead to unphysical non-perturbative non-linearities. 

\bibliography{mybiblio}

\end{document}